\documentclass[sn-mathphys,Numbered]{sn-jnl}

\usepackage{graphicx}%
\usepackage{multirow}%
\usepackage{amsmath,amssymb,amsfonts}%
\usepackage{amsthm}%
\usepackage{mathrsfs}%
\usepackage[title]{appendix}%
\usepackage{xcolor}%
\usepackage{textcomp}%
\usepackage{manyfoot}%
\usepackage{booktabs}%
\usepackage{algorithm}%
\usepackage{algorithmicx}%
\usepackage{algpseudocode}%
\usepackage{listings}%
\usepackage{caption}

\usepackage{amssymb,amsthm,latexsym,amsmath,mathrsfs,stmaryrd,graphicx}

\newcommand{\colc}[1]{\,{:}^{\sharp 1}\,}            
\newcommand{\colce}[1]{\,{:}^{e,{\sharp 1}}\,}       

\newcommand{\sem}[1]{\llbracket{\sharp 1}\rrbracket} 

\newcommand{\aware}[1]{E{\sharp 1}}                  

\newcommand{\rem}[1]{\relax}
\newcommand{\cut}[1]{\relax} 

\newcommand{\BB}{{\cal B}}

\newcommand{\Agents}{{\cal A}}

\newcommand{\abovearrow}[1]{\rightarrow\hspace{-.17in}\raisebox{1.0ex}
{$\scriptscriptstyle{#1}$}\hspace{.1in}}

\newcommand{\abovearrowlong}[1]{\longrightarrow\hspace{-.17in}\raisebox{1.0ex}
{$\scriptscriptstyle{#1}$}\hspace{.1in}}
\newcommand{\arrowA}{\,\lower1pt\hbox{$\abovearrow{A}$}}
\newcommand{\arrowa}{\,\lower1pt\hbox{$\abovearrow{a}$}}
\newcommand{\arrowAB}{\lower1pt\hbox{$\abovearrowlong{AB}$}}

\newcommand{\arrowBB}{\lower1pt\hbox{$\abovearrow{\BB}$}}
\newcommand{\arrowBBstar}{\lower1pt\hbox{$\abovearrowlong{\!\!\BB^*}\!\!$}}
\newcommand{\arrowAgents}{\lower1pt\hbox{$\abovearrow{\Agents}$}}
\newcommand{\arrowAgentsstar}{\lower1pt\hbox{$\abovearrow{\Agents^*}$}}
\newcommand{\arrowB}{\,\lower1pt\hbox{$\abovearrow{B}$}}
\newcommand{\arrowC}{\,\lower1pt\hbox{$\abovearrow{C}$}}
\newcommand{\arrowD}{\,\lower1pt\hbox{$\abovearrow{D}$}}
\newcommand{\arrowstar}{\,\lower1pt\hbox{$\abovearrow{*}$}}

\theoremstyle{thmstyleone}%
\newtheorem{theorem}{Theorem}
\newtheorem{proposition}[theorem]{Proposition}
\newtheorem{corollary}[theorem]{Corollary}

\theoremstyle{thmstyletwo}%

\theoremstyle{thmstylethree}%

\begin{document}

\title[Logic meets Wigner's Friend (and their Friends)]{Logic meets Wigner's Friend (and their Friends)}


\author*[1]{\fnm{Alexandru} \sur{Baltag}}\email{thealexandrubaltag@gmail.com}
\equalcont{These authors contributed equally to this work.}

\author*[2]{\fnm{Sonja} \sur{Smets}}\email{s.j.l.smets@uva.nl}
\equalcont{These authors contributed equally to this work.}


\affil[1,2]{\orgdiv{ILLC}, \orgname{University of Amsterdam}, \orgaddress{\street{Science Park 107}, \city{Amsterdam}, \postcode{1098 XG}, \country{the Netherlands}}}


\abstract{We take a fresh look at Wigner's Friend thought-experiment and some of its more recent variants and extensions, such as the Frauchiger-Renner (FR) Paradox. We discuss various solutions proposed in the literature, focusing on a few questions: {What is the correct epistemic interpretation of the
multiplicity of state assignments in these scenarios? Under which conditions can one include classical observers into the quantum state descriptions, in a way that is still compatible with traditional Quantum Mechanics? Under which conditions can one system be admitted as an additional `observer' from the perspective of another background observer? When can the standard axioms of multi-agent Epistemic Logic (that allow ``knowledge transfer'' between agents) be applied to quantum-physical observers? } In the last part of the paper, we propose a new answer to these questions, sketch a particular formal implementation of this answer, and apply it to obtain a principled solution to Wigner Friend-type paradoxes.}




\maketitle

\section{Introduction}\label{sec1}

In this paper we focus on Wigner's Friend thought-experiment \cite{Wi61}, and the proposed variations \cite{De85} and extensions such as the Frauchiger-Renner (FR) Paradox \cite{FrRe18}, that have recently shaken-up the debate in the foundations of quantum theory. Such thought experiments seem to indicate that, if quantum theory is assumed to be universally valid (and hence can be applied to complex systems\footnote{The notion of system is used to refer to any separated part of the universe that has a clear identity and to which the theory of Quantum Mechanics can be applied at a chosen level of abstraction that is meaningful for the system's identity. In this paper we will also use the term `system' instead of `complex system' when it is clear from the context.} that are composed of quantum systems as well as what would usually be described as `classical observers'), then different agents are rationally entitled to ascribe different (pure) states to the same system, and as a result they cannot share their information in a consistent manner. More precisely, the result in \cite{FrRe18} is stated in the format of a no-go theorem, stating that any theory which satisfies conditions (Q) (the universal validity of quantum theory), (C) (the consistency between agents about their predictions of measurement outcomes), (S) (the unique outcomes of measurements), will lead to a contradiction.

\medskip\par
The authors in \cite{FrRe18} conclude that one cannot consistently maintain all three conditions (S, C and Q), and hence that `quantum theory cannot be extrapolated to complex systems, at least not in a straightforward manner'. For some, the result seems to point towards a subjectivist interpretation of quantum mechanics, in which quantum states are ``unreal'' because of their observer-dependency. For others the solution lies in a relational interpretation, in which quantum reality simply consists of relative properties, that are always defined only with respect to some given `observing' system, but this does not in any way affect their ``real'', objective nature.

\medskip\par
According to the \emph{subjectivist (QBist) view}, a quantum state does not represent any objective fact of the world, or any true (factive) information that can be possessed by an agent (`knowledge'), but it simply captures the agent's subjective probabilistic beliefs about the possible outcomes of her measurements. These beliefs do not have any truth value, they are not right or wrong, they are just (in some sense) `rational' predictions with no objective content. Moreover, the different observers simply cannot incorporate each other's information in a consistent picture: they are condemned to ``agree to disagree'', on pain of contradiction. They seem to live in different universes, and no amount of communication can heal their disagreement!

\medskip\par
The \emph{relational view} \cite{Rovelli} is more nuanced, as it aims at keeping a form of \emph{realism}. State descriptions are now taken to reflect true `facts' of the world, though of a \emph{relative} nature: they are interactive properties, that a system $S$ possesses only in relation to another system $O$. Such binary, interactive properties are no less objective or `factual' than the classical unary properties. There is nothing inherently subjective about them, since the distinction between observed system and observer has nothing to do with subjectivity or beliefs, and it is just a methodological convention needed to describe the world from a \emph{situated} perspective; in Relational Quantum Mechanics, there is no ``view from nowhere''. But the distinction between $S$ and $O$ is arbitrary and the roles can be reversed: \emph{any} system can play the role of `observer'\footnote{The use of the notion of `observer' is in this paragraph taken to be in full alignment with the writing of C. Rovelli in \cite{Rovelli} where he states: ``By using the word ``observer'' I do not make any reference to conscious, animate, or computing, or in any other manner special, systems. I use the word ``observer'' in the sense in
which it is conventionally used in Galilean relativity when we say that an
object has a velocity ``with respect to a certain observer.'' The observer can be any physical object having a definite state of motion.''}, in the same way that any system of coordinates can play the role of a reference system in Relativity Theory.

\medskip\par
In another series of papers \cite{NdR,Boge,Corti}, authors analyze condition (C) in the language of modal logic so that it can be further specified in relation to (Q) and (S). The use of a precise logical language can indeed reveal insights into which parts of the paradox can lead to an inconsistency. Of course, if the set of assumptions one starts from are inconsistent then reasoning about them on the basis of a consistent set of axioms will still lead to a contradiction. In the case of the FR paradox scenario, it is clear that one can reason about each of the assumptions (C),(Q) and (S) separately in a consistent way while the problem appears when we merge (C),(Q) and (S) and their underlying axiom systems together. None of the logic-based approaches in the literature actually show what an adequate constraint on the merged axiom systems should be to obtain a consistent set of axioms under which we can reason about all assumptions together. Simply denying the validity of the basic axioms of Epistemic Logic (as suggested in \cite{NdR}), just because they imply condition (C), seems to us a rather ad hoc and unhelpful approach, since it begs the question: what other epistemic principles are we supposed to use when talking about quantum agents? And why do the standard postulates of Epistemic Logic \emph{seem} to work for reasoning about practically all multi-agent protocols used in (classical or quantum) computation and communication?

\bigskip\par
In this paper we assume the standard Hilbert-space formalism of quantum theory (as formalized by von Neumann in \cite{VN}), as well as the common-sense use of higher-order level reasoning of classical agents about each other (as formalized in basic Epistemic Logic \cite{Halpern}). But we aim to provide a clear-cut criterion for \emph{when is such epistemic reasoning applicable to `agents' that are quantum-physical systems}. This requires us to address some related questions:
first, what is the correct epistemic interpretation of the multiplicity of state assignments? And secondly we ask under which conditions can one actually include classical observers in the quantum state descriptions in a way that is still compatible with traditional Quantum Mechanics?  As we show in this paper, once these questions are answered, the issue of communication and agreement between different observers can be easily addressed.

\medskip\par
To answer these questions, we will need to clarify a topic that the standard formalism of quantum mechanics has left open: what is a precise specification of what counts as an `observer' and what counts as `an observed system' in scenarios where we have more than 1 observer?\footnote{Note that von Neumann ends his work in \cite{VN} with the remark that the discussion of more complicated examples which include `the control that a second observer might effect upon the measurement' can be carried out in the same fashion as he has done for one observer but he stops here and leaves it to the reader to work this out.} In the case of modelling the quantum observation performed by a single classical observer, both Heisenberg \cite{Crull} and von Neumann \cite{VN} have indicated that it is necessary to start by specifying a clearly dividing line, a \emph{cut} $c_{SO}$ between what counts as the system $S$ under observation and what counts as the `observer' $O$.
The exact place of the cut has been a topic of debate in the foundations of quantum theory, it is a problem that has been referred to as the ``shifty split'' problem by J.S. Bell. Von Neumann specifies that the cut $c_{SO}$ can, at least to a very large extent, be \emph{arbitrarily chosen}, and hence that the system under observation could well include the measurement apparatus or even the chemical processes that happen in one's brain. Similarly, Heisenberg indicates that the cut is movable and can be ``shifted arbitrarily far in the direction of the observer in the region that is otherwise described according to the laws of classical physics'' \cite{Crull}. The question Heisenberg is concerned with is to show that ``the quantum mechanical predictions about the outcome of an arbitrary experiment are independent of the location of the cut...'' \cite{Crull}. The use of the term `cut' in this context plays a double role for both von Neumann and Heisenberg, on the one hand the cut is used to indicate which parts of reality are called $S$ and $O$ and on the other hand it is used to indicate which parts of reality are described by quantum theory and which parts fall out of that description.

The notion of observer that von Neumann and Heisenberg adopt seems to work fine as long as we consider only one observer, but Wigner's Friend scenarios require us to reason about the cuts that are tied to different observers. How we should reason about another observer's measurement results? In other words, given a specific cut $c_{SO}$, when is an observer $O$ entitled to admit some (other part of the) observed system $S$ as an \emph{additional observer} $O'$ (who comes with her own cut $c'_{O'S'}$), in a way that is still consistent with the predictions made by standard quantum theory?\footnote{We refer to \cite{NuRe21} for an overview of how different interpretations of quantum mechanics treat the cut between an observer and a system under observation, where in particular the neo-Copenhagen interpretation can be aligned with the view that takes the cut to be subjective so that each observer induces her own cut.} Without first answering this question, it doesn't even make sense to talk about higher-order reasoning (of observers about other observers): such a specification (of when another physical system can count as an observer for us) needs to be first checked \emph{before} we can proceed to reason about other systems \emph{as observers}.

\medskip\par
The solution we propose in this paper can be understood as extending the relational view on Quantum Mechanics to the property of ``being an observer'', which becomes itself relativized. More specifically, we define a relational notion of \emph{admissible observer} (that is always relative both to another background observer $O$ and to a given history $h$ or protocol $\pi$). Roughly speaking, a system $A$ is an ``admissible observer'' with respect to $O$ and $h$ if and only if it is always possible that (as far as the background observer $O$ can know) that none of the information carried by $A$ will be fully erased from the universe at any moment of the given history $h$.\footnote{The notion of admissible observer is applicable to any system, which includes a measurement apparatus that only records a certain value as well as a human being (or AI) who perceives the outcome and can communicate and reason about it. Concrete, our use of `admissible observer' includes physical systems but also systems that are typically referred to as `agents' in epistemic logic, AI and computer science. As such the idea of the `cut' is used to demarcate the difference between $O$ and $S$ but we no longer identify this split with what can and cannot be described by quantum theory.} This can happen either because $A$'s memory stays intact until the end of $h$, or because the information keeps ``leaking'' out of $A$, e.g. in the form of copies or records that are being disseminated and copied again before they can be destroyed.\footnote{This second case is actually the most general one, since it subsumes the first case: $A$ can in a trivial sense be said to be a record of itself.}

A further step is to require \emph{mutuality} of this relationship: a family of subsystems forms a ``community of admissible observers'' if each of them is an admissible observer with respect to all the others. Full higher-order epistemic reasoning (of observers reasoning about other observers etc.) is consistently applicable only within such communities of admissible observers.

Typical Wigner's Friend-type scenarios break these conditions: some of the supposed ``agents'' (such as the Friend in the box) are \emph{not} admissible observers from the perspective of the other agents (such as Wigner), precisely because the later ones get to \emph{know} that the information carried by the earlier one is being completely erased from the universe. In contrast, in the epistemic scenarios that underly standard multi-agent protocols in Quantum Information and Quantum Computation theory, one can claim that the conditions required for mutual admissibility as observers \emph{are} satisfied.\footnote{ This is primarily because large macroscopic systems are inherently ``leaky'', spontaneously disseminating information into their environment. And secondarily, because real `agents' typically do make multiple copies of their information, and tend to save them or leak them into places that are not under the full control of other `agents' (and may be even beyond their own control). It is thus practically impossible for an agent to be absolute sure that all traces of another agent's information have been fully erased from the universe.}

\bigskip\par\noindent
We organize this paper as follows. In the next section, we first introduce Wigner's Friend paradox and its extensions, and in section \ref{subsec22} we briefly sketch how in certain conditions the paradoxes can be avoided by allowing agents to ``leak'' their information into remote places (or by simply keeping their self-record intact) that are beyond the other agents' control. In subsection \ref{subsec23} we look at the stronger FR Paradox, and in subsection \ref{subsec24} we go over some of the solutions and interpretations of this thought experiment proposed in the literature. In section \ref{sec3}, we sketch a formalization of our proposed solution and we apply it to several versions of the above-mentioned paradoxes. We end, in section \ref{sec4}, with some conclusions and further reflections.

\section{Introducing Wigner's Friend(s)}\label{sec2}

Wigner's Friend thought experiment \cite{Wi61}, as well as the proposed variations and extension by Deutsch \cite{De85,Brukner}
and in particular the \emph{Frauchiger-Renner (FR) Paradox } \cite{FrRe18}, aim to show that if quantum theory is assumed to be universally valid (and hence can be applied to complex systems that are composed of quantum systems as well as their classical observers), then different agents are rationally entitled to ascribe different (pure) states to the same system, and as a result they cannot share their information in a consistent manner.
\medskip\par
The latest result in this series is the FR-paradox which is stated in the format of a no-go theorem, stating that any theory which satisfies conditions (Q) (the universal validity of quantum theory), (C) (the consistency between
agents about their predictions of measurement outcomes), (S) (the unique outcomes of measurements), will lead to a
contradiction. It is crucial that these assumptions are made fully explicit, so we will provide our reading of them as follows.

\medskip\par\noindent
\textbf{Assumption (S)}: Quantum mechanics in von Neumann's formalisation describes external systems $S$ from the perspective of an observer $O$. When $O$ interacts with $S$, he describes this as a state collapse, with a unique measurement result. According to von Neumann, ``we are obliged always to divide the world into two parts, the one being the observed system, the other the observer. In the former we can follow all physical processes (in principle at least) arbitrarily precisely. In the latter, this is meaningless. The boundary between the two is arbitrary to a very large extent.''\cite{VN}. In this view, the ``cut'' between the observer $O$ and the observed system is here taken to be relative and irrelevant. Indeed every system, be it a measurement apparatus or the retina of an observer, i.e. a system that can record a result, can be treated as an `observer'.

\medskip\par\noindent
\textbf{Assumption (Q)}: This assumption captures `the universal validity of quantum theory'. Following \cite{VN}, given an external system $S$ and observer $O$, then, in the absence of interaction with $O$, $O$ can describe $S$ using Quantum Theory. More specifically, she can use unitary evolutions $U$ (possibly applied to a larger supersystem $S'\supseteq S$, obtained by taking a tensor product), then possibly applying a partial trace (to represent the state of subsystem $S$).
\smallskip\par\noindent
In the case there is an interaction between $O$ and $S$, then assumption (Q) indicates that by adopting the `external' perspective of another observer $O'$ (that is far enough from any given interaction between $O$ and $S$), every 'apparent' measurement-collapse can also be treated as a unitary evolution of the combined system $O+S$ (sometimes combined with taking a partial trace, to calculate only the state of system $S$, which is typically given by a mixed state in this case). E.g. in the context of the scenario of the FR-paradox, the authors of \cite{FrRe18} stress that (Q) implies on the one hand that the Born rule is a universal law, that can be used by $O$ to predict the outcome to a measurement on any system $S$ that she is interacting with (-even if $S$ includes other observers, but not $O$ herself). On the other hand, in our view, (Q) also implies that, from the perspective of an external observer $O'$ (that is outside $S+O$), the same interaction between $O$ and $S$ can be described as a unitary evolution. We believe that our reading of (Q) is fully aligned with how quantum theory is actually applied in the context of these paradoxes.
\medskip\par\noindent
\textbf{Assumption (C)}: The descriptions obtained by using the standard formalism of quantum mechanics by different `observers', are mutually consistent in some sense, and can thus be shared (by communication) without contradictions. Even in the absence of communication, observers can still put themselves in the shoes of other observers, reason counterfactually (including about themselves) from such an external perspective, and in this process use freely their own current private information, with no restrictions and no danger of contradictions.

\medskip\par
In this explicit form and under these assumptions, the ``FR paradox'' is just a \emph{theorem}, there is no arguing with it. So the remaining questions we are concerned with are: (a) which assumption(s) are subject to constraints or must be given up, and why? (b) how can we explain the apparent validity of all the above assumptions in ``standard'' contexts (e.g. quantum information protocols involving multiple agents, the non-contradictory nature of science as a social activity based on communication and agreement of multiple scientists, etc.)?

\subsection{Wigner's Friend thought experiment}\label{subsec21}

The original version of Wigner's Friend paradox (as depicted in Fig. 1) assumes as given an isolated lab $L$, within which there is a quantum system $S$ and a friend $F$, who can perform measurements on $S$. Wigner $W$ is located outside the lab where he can perform measurements on the entire lab $L$ consisting of $S$ and $F$. Both $F$ and $W$ are treated as different `observers'. The following scenario ensues.

\begin{figure}[h]
    \centering
       \label{fig:some_photo}
    {\scalebox{0.25}{\includegraphics[width=204mm]{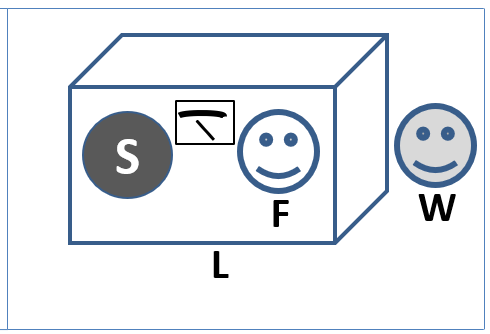}}}
\caption{{\small{Wigner's Friend (F) measures system $S$ inside the lab, while Wigner (W) is positioned outside and measures the entire lab $L$ consisting of the system $S$ and the friend $F$.}}}
 \label{fig:Fig1}
 \end{figure}

\smallskip\par\noindent
{\bf Epistemic-Quantum Scenario 1.} The experiment starts by assuming that it is common knowledge\footnote{Common knowledge is one of the strongest epistemic requirements that one can impose on the members of a group, it requires an infinite series of higher-order levels of knowledge about a fact to hold (i.e. $p$ holds and everybody knows it and everybody knows that everybody knows it and everybody knows that everybody knows that everybody knows it etc.). For details see a standard textbook such as \cite{Halpern}.} that a quantum system $S$ is initially in the superposition state $|+\rangle_S:= \frac{1}{\sqrt{2}} (|0\rangle_S+ |1\rangle_S)$. Inside the \emph{Lab} $L=S+F$, the \emph{Friend} $F$ measures system $S$ in the standard basis $\{|0\rangle_S,|1\rangle_S\}$, recording either the outcome $a=0$ for state $|0\rangle_S$ or $a=1$ for state $|1\rangle_S$. Suppose the actual outcome is $a=0$. Looking now at Wigner's perspective, $W$ describes the entire lab-experiment as a unitary transformation $U$, that entangles $S$ and $F$. When setting $|fail\rangle_L:=\frac{1}{\sqrt{2}}(|0\rangle_S\otimes|0\rangle_F + |1\rangle_S \otimes |1\rangle_F)$, and
$|ok\rangle_L:=\frac{1}{\sqrt{2}}(|0\rangle_S\otimes|0\rangle_F - |1\rangle_S \otimes |1\rangle_F)$, $W$ computes the result of $U$ to be  $|fail\rangle_L$.

\smallskip\par\noindent
{\bf Reasoning from different perspectives.} The problem in scenario 1 becomes visible by the observers assigning different descriptions to the state of $S$ and providing different predictions. Indeed at the end of the scenario, $W$ and $F$ have the following \textit{different descriptions} of the system $S$: $F$ assigns to $S$ the pure state $|0\rangle_S$; while $W$ assigns to $S$ the mixed state $\{|0\rangle_S: \frac{1}{2}, |1\rangle_S: \frac{1}{2}\}$, obtained by tracing out $F$ in the density-operator description of $L$:
$$Tr_F (\rho_L):= Tr_F (|fail\rangle_L \langle fail|_L)= \frac{1}{2} |0\rangle_S \langle 0|_S +\frac{1}{2} |1\rangle_S\langle 1|_S.$$

\smallskip\par\noindent
These different descriptions by $F$ and $W$ lead to \textit{different predictions}: $F$ predicts that any new measurement of $S$ in the standard basis will yield outcome $|0\rangle_S$; while $W$ assigns equal probabilities to the outcomes $|0\rangle_S$ and $|1\rangle_S$.
Whose predictions are `right': $F$'s or $W$'s?

\bigskip\par\noindent
{\bf Compatible State Descriptions.} First we note that, in a sense, the descriptions of $F$ and $W$ at the end of scenario 1 are {\it compatible} with each other, as one is a (more informative) \emph{refinement} of the other. Indeed, the probabilistic assignment $\{|0\rangle_S: 1\}$ can be obtained from the assignment $\{|0\rangle_S: \frac{1}{2}, |1\rangle_S: \frac{1}{2}\}$ by applying standard \textit{\emph{Bayesian conditioning}}.\footnote{The use of probabilistic methods such as Bayesian conditioning does not necessarily imply that the quantum state descriptions by observers have to be identified with a set of probabilistic credences or that the quantum state needs to be given a subjective interpretation.}

\smallskip\par
More generally, Leifer and Spekkens \cite{LF} define two density operators (``mixed states'') to be \textit{compatible} if they have a \emph{common refinement} (obtainable by conditioning each of the two assignments); equivalently, iff they can both be obtained (by conditioning) from a \emph{common prior} assignment. In particular, density operators corresponding to pure states (``maximally informative'' descriptions) are compatible iff they are equal.

\smallskip\par
Since the probabilistic assignment $\{|0\rangle_S: 1\}$ is a refinement of the assignment $\{|0\rangle_S: \frac{1}{2}, |1\rangle_S: \frac{1}{2}\}$, we can conclude that $F$'s description is \emph{more informative} than $W$'s in scenario 1. Hence in a sense both $W$ and $F$ can be said to be ``right'', and agreement is possible, even though $F$ has more information about $S$. This claim can be ``confirmed'' by communication: suppose $F$ announces to $W$ the outcome $a=0$ of his measurement. From $W$'s perspective, this can be interpreted as a measurement by him of $F$'s state $|0\rangle_F$, which collapses the state
 $|fail\rangle_L=|0\rangle_S\otimes|0\rangle_F + |1\rangle_S \otimes |1\rangle_F$ into the state $|0\rangle_S\otimes |0\rangle_F$. The two agree now, and $W$ has now adopted $F$'s description of $S$.

\smallskip\par
The above argument seems to indicate that \textit{$F$ \emph{really} has more information about $S$ than $W$}! After all if $F$ were to announce his measurement outcome to $W$ then they would agree. But is this argument always applicable?

\bigskip\par\noindent
{\bf Incompatible State Descriptions.} In the above analysis of scenario 1 we pointed out that $F$'s description is more informative than $W$'s description, if we are concerned with the question: what will be the outcome of a measurement of $S$ in the standard basis?
But what if we are concerned with predicting the outcome of measurements of the full lab $L=S+F$? In that case we reach a \textit{problem} because $F$ \textit{cannot even represent} a measurement of $L$ which includes himself.
In general, Quantum Mechanics \emph{does not seem to provide a way for observers to describe themselves} (or any supersystem that includes themselves).
\smallskip\par
To still reason about the lab $L$ that contains himself, agent $F$ could try to adopt an ``external'' (counterfactual) view of himself, by asking himself what state $W$ (or any other external observer) would assign to the lab $L$ if he ($F$) communicated to him all he knows?
As we saw, after such a communication, $W$ would describe $L$ as being in state $|0\rangle_S\otimes |0\rangle_F$. The same applies to any external observer who'd have access to the information of both $W$ and $F$.
Note that, to do this, \emph{no actual communication is necessary}: $F$ can just imagine this communication as a counterfactual possibility, and conclude that the state of his lab $L$ is $|0\rangle_S\otimes |0\rangle_F$.
Since no communication actually happens, $W$ still assigns the state $|fail\rangle_L$ to the lab $L$.

\smallskip\par
We seem to have reached a contradiction. The two descriptions $|0\rangle_S\otimes |0\rangle_F$ (as provided by $F$ when he reasons from $W$'s perspective and takes his own information into account) and $|fail\rangle_L$ (as provided by $W$)  are both \emph{pure states}, thus they are \textit{incompatible} in the sense of being mutually exclusive. The obtained descriptions cannot both represent the true state, since they do not have a common refinement, and they lead to incompatible predictions.

\medskip\par
One could in fact claim that in some sense \emph{each of the agents has `more' information than the other with respect to \emph{some} potential measurement}.
To recap, if the lab $L$ is measured in the standard basis, then $F$'s description predicts outcome $|00\rangle_L$ with probability $1$ while
$W$'s description gives equal probabilities $\frac{1}{2}$ to outcomes $|00\rangle_L$ and $|11\rangle_L$. So $F$ has more information than $W$ about standard-basis measurements of $L$.
However, if the lab $L$ is measured in any basis that includes $|fail\rangle_L$ and $|ok\rangle_L$, then $W$'s description predicts outcome $|fail\rangle_L$ with probability $1$; while $F$ gives  equal probabilities $\frac{1}{2}$ to outcomes $|fail\rangle_L$ and $|ok\rangle_L$.
So one could claim that $W$ has more information than $F$ about the Bell-basis measurements of $L$.

\smallskip\par
So, who is `right'? Can both agents be right? This time, quantum mechanics provides no way to combine the information of the two observers: there is no common refinement to any (pure or mixed) state.

\bigskip\par\noindent
{\bf Reasoning about Other Observers in Quantum Mechanics.} The lesson we draw from the analysis of scenario 1 is that reasoning about other agents is problematic in quantum mechanics. It seems that if $F$ adopts an `external' perspective on himself while combining this view with his own information, then the two agents' descriptions are \emph{directly contradictory}!

\smallskip\par
But note that \emph{this contradiction is not ``centralized'' in any one agent}: no {single} observer possesses all the information needed to derive the contradiction. Each of the descriptions is still ``locally consistent''. Also, note that an actual measurement by $W$ (or anybody else) of the lab $L$ in a basis that includes $|fail\rangle_L$ and $|ok\rangle_L$ would \emph{erase the quantum state} $|0\rangle_S$, thus \textit{physically destroying the information that $F$ possessed about $S$}. On the other hand, we already saw that, if instead of this, the friend $F$ communicates his information to $W$, then $W$'s assignment of $|fail\rangle_L$ to the lab changes to $|00\rangle_L$. In both cases, $W$ and $F$ end up agreeing (one way or another) after one's or the other's information flow between them (either by a destructive measurement by $W$ or by a communication by $F$ of the results of his measurement).
\smallskip\par\noindent

\smallskip\par
So maybe a solution to the paradoxes could be to simply admit that observers may have incompatible perspectives, \emph{as long as the resulting global inconsistency is never ``internalized'' in any single agent}? This would reconciliate the relativity of perspectives with the consistency of each local perspective, as well as with observed agreement between communicating observers. However, this easy solution still runs into problems, since we can restate the paradox in terms of \emph{two opposing perspectives that can be adopted by \emph{the same} observer}.

\bigskip\par\noindent
\textbf{Internalizing the contradiction.} Reasoning only from the perspective of Wigner, he can still adopt two different perspectives: (1) according to $W$'s ``\emph{Minimal Cut}'', $W$ considers himself as the \emph{only collapsing observer}: all other interactions are modeled as unitary evolutions. He assigns to $L$ the \emph{pure state}
$|fail\rangle_L$. But on the other hand: (2) $W$ can alternatively
adopt an ``\emph{Extended Cut}'', 
admitting $F$ as an \emph{additional collapsing observer} (although \emph{one whose measurement results are not automatically known to $W$}).
Now he assigns to $L$ the \emph{mixed state} 
$\{|00\rangle_L: \frac{1}{2}, |11\rangle_L: \frac{1}{2}\}$, given by the density operator
$\frac{1}{2} |00\rangle_L \langle 00|_L +\frac{1}{2} |11\rangle_L\langle 11|_L$. Quantum mechanics says that the cut is `arbitrary', so in principle \emph{this choice shouldn't matter} (as long as it includes the observer $W$). Yet it \emph{does matter}: the two states \emph{yield different predictions}!

\bigskip\par\noindent
So, when reasoning about measurements of the lab $L$, we seem to have reached a genuine contradiction! It looks like quantum mechanics doesn't allow observers to adopt another observer's perspective. Or should we just conclude that agents simply cannot reason about actions that erase their own current information? Or maybe they can do that, as long as they abstain from using the erased information? Or maybe even that is OK, as long as the resulting global inconsistency is never ``internalized'' in any single agent? We come back to such questions in the next sections.

\subsection{The Third Observer and the Leaking Lab}\label{subsec22}

One can try to avoid the issue of agents' reasoning about actions that erase their own memory, by delegating this task to a third observer $O$.
As we'll see below, an agreement about the state-descriptions can be regained if $F$ ``leaks'' his information out of the lab before it is erased by $W$, by communicating it to $O$. To show this we introduce scenario 2.
\bigskip\par\noindent
{\bf Scenario 2: Wigner's Friend and an Observer.} In this altered scenario, we assume that $O$ starts in a state $|0\rangle_O$. Initially, $O$ is in the same epistemic situation as $W$ is in scenario 1 : so $O$ assigns to $L$ the state $|fail\rangle_L$.
\smallskip\par\noindent
At this stage, we have three observers $W$, $F$ and $O$: the two external observers ($W$ and $O$) agree on the lab's state assignment $|fail\rangle_L$, but their state description is incompatible with $F$'s assignment of state $|00\rangle_L:=|0\rangle_S\otimes |0\rangle_F$ to the same lab. These different predictions could in principle be tested against each other by performing a measurement of the lab $L$ in the standard basis (or also in the Bell basis).

However, suppose now that, \emph{before} any such measurement of the whole lab $L$ can be performed, $F$ communicates the outcome of his measurement $a=0$ to $O$. When $O$ receives the message from $F$, the state assigned to $L$ by $O$ collapses to $|00\rangle_L$. Now $O$ and $F$ agree on their descriptions of the lab (and hence also on their descriptions of system $S$): they both assign the state $|00\rangle_L$ to the lab $L$, and state $|0\rangle_S$ to system $S$. How about Wigner?

We might think that $W$ (being unaware and unaffected by the communication between $F$ and $O$) will still assign to $L$ the state $|fail\rangle_L$.
\smallskip\par\noindent
In scenario 2, it seems that we still have obtained a direct contradiction between the descriptions of $L$ by $W$ and $O$ while having avoided any reasoning about one's memory erasure, or any adoption of another agent's perspective!

\smallskip\par\noindent
Note however that, given scenario 2, this reasoning is actually \textit{mistaken}: it relies on $W$'s \emph{wrong assumption} that no communication happened between $F$ and $O$. If $W$ assigns to the lab the state assignment $|fail\rangle_L$, this simply reflects his \emph{false belief} that $F$ has not leaked his measurement results to anybody else.

However, if $W$ is a perfectly rational agent, he cannot rely on this false assumption. Even the \emph{mere possibility} of such a communication should lead $W$ to \emph{change} his state assignment for $L$. Indeed, let us suppose that $W$ considers possible that $F$ is leaking his results to $O$, but assigns a very low (though non-zero) probability to this eventuality, say, $1\%$ (probability $0.01$); hence, with probability $0.99$, he thinks no leaking happens. In the case that the communication does happen, $W$ should model this as a unitary operation that entangles $L$ and $O$, ending with a description of system $L+O$ in the state
$$|fail\rangle_{LO}:=\frac{1}{\sqrt{2}}(|00\rangle_L \otimes |0\rangle_O + |11\rangle_L\otimes |1\rangle_O).$$
In contrast, in the case that no leaking happens, then $W$ should still assign state $|fail\rangle_L$ to the lab (and thus assign state $|fail\rangle_L\otimes |0\rangle_O$ to the system $L+O$). Overall, given his probabilistic uncertainty about these two cases, $W$ should assign to $L+O$ the mixed state
$$\{ |fail\rangle_{LO} : \frac{1}{100}, |fail\rangle_L\otimes |0\rangle_O : \frac{99}{100} \}.$$
By tracing out $O$'s state (i.e. taking the partial trace $Tr_O$), we conclude that $W$'s correct description of the lab is given by the mixed state
$$\{|00\rangle_L :\frac{1}{200}, |11\rangle_L: \frac{1}{200}, |fail\rangle_L: \frac{99}{100}\}.$$
This description is \textit{compatible} with both $O$'s and $F$'s descriptions of $L$ as $|00\rangle_L$, it is indeed a less informative description (that can be refined by Bayesian conditioning to $|00\rangle_L$). Now, all agents agree in their predictions concerning any further measurement of the lab (in the sense that their state descriptions are compatible)!

If instead we make the stronger assumption that $W$ \emph{knows} that the leaking happens (without knowing the communication's actual content, i.e. $F$'s measurement results), then his description gets sharper: since he assigns probability $1$ to the first case (in which communication happens), his state assignment for $L+O$ will now be exactly  $|fail\rangle_{LO}$, and thus by tracing out $O$ he will assign to the lab the mixed state
$$\{|00\rangle_L :\frac{1}{2}, |11\rangle_L: \frac{1}{2}\}.$$

\smallskip\par\noindent
So we see that, if we allow for the possibility (or alternative, for the certainty) of information `leaking' out of the lab (before it is erased by $W$ and before the lab as a whole is measured), agreement (or rather, compatibility of state descriptions) is regained!
As far as $W$'s description of $L$ is concerned, both the above mixed-state assignments for $L$ are compatible with the description that $W$ would give to $L$ if he included $F$ (together with himself) on the observer's side of the `cut' (i.e. if he assumed that, whenever $F$ interacts with a system $S$, it collapses its state, without $W$ necessarily knowing the actual outcome). Indeed, $W$'s description of $L$ in that case would be
$$\{|00\rangle_L :\frac{1}{2}, |11\rangle_L: \frac{1}{2}\},$$
which coincides with his description above in the case that leakage is certain (and is a refinement of his description for the case that leakage is merely possible with some non-zero probability).

\medskip

It is true that theoretically this agreement is not `perfect': just before the leakage, there is disagreement between the internal observer ($F$) and the external ones ($W$ and $O$) concerning $L$'s state. But, if any measurement of the whole lab $L$ can only happen after the leakage happened (with certainty), or at least after it may have happened (as far as the relevant agents know), then this disagreement remains purely `academic' and in this case also essentially untestable: \emph{the possibility of information leaking saves the appearances}. To put it conversely: if the friend $F$'s information is inherently so ``leaky'', or if he is inherently so prone to making records of his information and disseminating them far enough beyond $W$'s control, then for all practical purposes $W$ \emph{can} treat $F$ as a collapsing `observer', just like himself!

\medskip

One may still object that, even after the leaking happens, there will still be some disagreement at some higher level, when the different observers attempt to describe the super-system $L+O$. E.g. in the case that $W$ knows that the communication between $F$ and $O$ happened (but doesn't know $F$'s measurement result), we saw that he assigns state $|fail\rangle_{LO}$ to $L+O$. In contrast, at the same time (after the communication from $F$ to $O$), both $F$ and $O$ know the result of $F$'s measurement was $|0\rangle_S$ (and know that $O$ got this result); so (by counterfactually taking the point of view of some other observer external from themselves) they could describe the super-system $L+O$ as being in state  $|fail\rangle_L\otimes |0\rangle_O$. We again obtain incompatible predictions, this time referring to the results of possible measurements of $L+O$. However, this disagreement also remains `academic' and in this case also untestable, as long as no measurement of the whole $L+O$ can happen, e.g. because this super-system cannot be encapsulated, but remains open and beyond our agents' control; which means that, even if and when such a global measurement of $L+O$ would eventually happen, further leaking (outside $L+O$) may have happened in the meantime (or at least this possibility cannot be excluded by them).

\subsection{FR thought experiment}\label{subsec23}

As we saw, all the above paradoxes involve the physical erasure of the information possessed by one of the agents involved in each scenario. As we also noticed (and as previously noted by other authors \cite{WvN,Laloe}), these scenarios stop being `paradoxical' if, before any such erasure takes place, the relevant information has been copied, recorded or ``leaked'' in some remote place (that is not under other agents' control). In fact, as we will show, the mere \emph{epistemic possibility} of such leakage is enough to solve the paradox: as long as nobody knows for a fact that the relevant information has been completely erased from the universe, no paradox will occur! In section \ref{sec3}, we will take this observation as the basis of our proposed solution of the paradoxes.

But before doing this, we first have to consider a version of these paradoxes that may at first sight seem to provide a pre-emptive objection to this solution: what if we manage to get the contradiction, while apparently still preserving the observed information somewhere else, e.g. into another observer who keeps a record on it? This is exactly the claim involved in
the recent strengthening of Wigner's Friend proposed by Frauchiger and Renner (the \emph{FR Paradox}) \cite{FrRe18}.
In addition, what if at the same time we avoid any need for internalizing another agent's perspective, by reasoning strictly from one agent's perspective (as in \cite{Healey})?

\smallskip\par
The FR thought experiment, and its subsequent variants, were designed precisely in order to make explicit these issues and explore their puzzling implications.

\begin{figure}[h]
    \centering
  \scalebox{0.5}{\includegraphics[width=204mm]{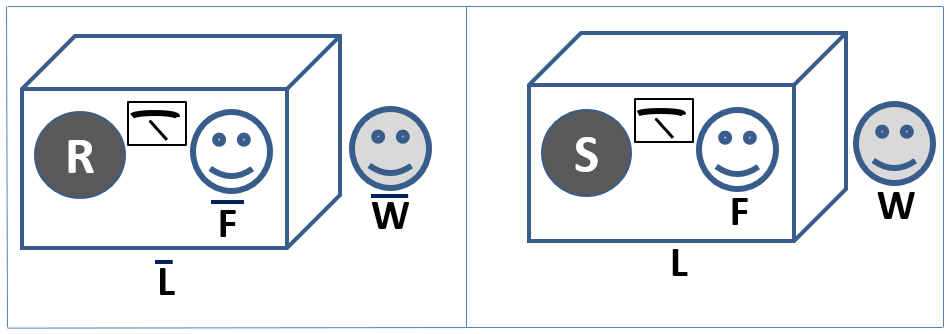}}
    \caption{{\small{The FR thought experiment consisting of 4 agents (F,$\overline{F}$,W,$\overline{W}$) and two labs, $L$ and $\overline{L}$. In the first lab $\overline{L}$, agent $\overline{F}$ can measure system $R$ while agent $\overline{W}$ can measure the lab $\overline{L}$. The preparation state of system $S$ will depend on the state of system $R$ after $\overline{F}$ has measured it. In the second lab $L$, agent $F$ can measure system $S$ while agent $W$ can measure the lab $L$. The agents can perform communication actions.}}}
    \label{fig:Fig.2}
\end{figure}


\par\noindent
{\bf Scenario 3 (Frauchiger and Renner 2018).}  This time we have two labs $L$ and $\overline{L}$, two outside observers (two `Wigners' $W$ and $\overline{W}$), and two `Friends' ($F$ and $\overline{F}$) inside their corresponding labs (see Fig.2). In addition, each lab will contain an observed subsystem: a quantum coin $R$ inside $\overline{L}$, and a particle spin $S$ (to be sent as a signal by $\overline{F}$ and received by $F$ inside $L$).

The protocol starts with the following events in lab $\overline{L}$:

\begin{itemize}

\item At {$t_0$}: $\overline{F}$ measures a quantum coin $R$ in state $\sqrt{1/3}|h\rangle_R + \sqrt{2/3}|t\rangle_R$ and measures it in the $\{|h\rangle_R,|t\rangle_R\}$ basis. $F$ outputs $r=heads$ or $r=tails$ with probabilities $1/3$ and $2/3$ respectively,  inside lab $\overline{L}$.
\item Still at {$t_0$}: $\overline{F}$ sets the spin $S$ of a particle to $|\downarrow\rangle_S$ if $r=heads$ and to superposition state $|\rightarrow\rangle_S$ = $ \sqrt{1/2}(|\downarrow\rangle_S + |\uparrow\rangle_S)$ if $r=tails$. Next, she sends $S$ to $F$.
\end{itemize}


The protocol continues with the following events in lab $L$:

\begin{itemize}

\item At {$t_1$}: $F$ measures $S$ in the basis $\{|\downarrow\rangle_S,|\uparrow\rangle_S\}$, records the outcome.

    \end{itemize}

Next we return to the actions of the outside observers, we start first with lab $\overline{L}$:


\begin{itemize}
\item At {$t_2$}: $\overline{W}$ measures the lab $\overline{L}=\{R,\overline{F}\}$ w.r.t the basis containing $|\overline{ok}\rangle_{\overline{L}}$
    with \par\noindent
   $$|\overline{ok}\rangle_{\overline{L}}=
    \sqrt{1/2}(|h\rangle_{R}|h\rangle_{\overline{F}} - |t\rangle_{R}|t\rangle_{\overline{F}})$$   
   She announces $\overline{ok}$ if the outcome associated to this vector occurs else, she announces $\overline{fail}$.
\end{itemize}

The protocol continues with $W$ observing the second lab $L$:


\begin{itemize}
\item At {$t_3$}: $W$ measures lab $L=\{S,F\}$ w.r.t a basis containing  $|ok\rangle_L$,
    where
    $$|ok\rangle_L = \sqrt{1/2}(|\downarrow \rangle_{S}|\downarrow\rangle_F - |\uparrow\rangle_{S}|\uparrow\rangle_F)$$
            She publicly announces $ok$ if the outcome associated to
            this vector occurs, else she announces $fail$.

\end{itemize}
If $\overline{W}$ gets outcome $\overline{ok}$ {\it and} $W$ gets $ok$ then the experiment is halted, else the protocol repeats.

\smallskip\par\noindent
From the modeler's perspective, when using quantum mechanics, the halting condition is reached with probability $1/12$. However, epistemic reasoning indicates that the halting condition can never be reached, and hence we obtain a direct contradiction: from $W$'s perspective, if he hears $\overline{W}$ announcing `$\overline{ok}$' then he ($W$) is certain that in the final step he'll get `$fail$' (and thus the pair of outcomes $(\overline{ok}, ok)$ is never reached). The proof, as described in more detail by Waaijer and van Neerven \cite{WvN}, goes via the following epistemic reasoning steps:
\begin{itemize}[rightmargin=*]
\item[(i)] If $\overline{F}$ observes `tail' at $t_0$, she prepares $S$ in the superposition state and sends it to $F$. She also infers that $W$ will announce `fail'.
\item[(ii)] If $F$ measures `up' at $t_1$ , she is certain that $\overline{F}$ must have measured `tail' at $t_0$. Because of (i) and the  hypothesis (C), $F$ therefore is certain that $W$ will announce `fail'.
\item[(iii)] If $\overline{W}$ measures `$\overline{ok}$' at $t_2$, he infers that $F$ must have measured the spin to be `up' at $t_1$ . Because of (ii), $\overline{W}$ is then certain that $F$ is certain that $W$ will
announce `fail'. Hence by (C), $\overline{W}$ is certain that $W$ will announce `fail'.
\item[(iv)] If $W$ hears $\overline{W}$'s announcement of `$\overline{ok}$' at $t_2$, then by (iii) $W$ is certain that $\overline{W}$ is certain that $W$ will announce `fail'. Hence by (C), $W$ is certain to announce `fail'.
\end{itemize}

\subsection{Approaches to the FR Paradox}\label{subsec24}

Several authors (see e.g. \cite{FrRe18,NdR,WvN,DeBrota,CR,Brukner}) conclude that quantum mechanics \emph{cannot consistently describe one observer's reasoning about another observer}, at least not in the way in which common sense (and standard epistemic logic) does. Hence, some of these authors question condition (C), i.e. the assumption that the agents' descriptions/predictions are mutually \emph{consistent}. These doubts can be extended to one or another of the formal conditions underlying (C), depending on which of them are deemed blameworthy by each author.

\medskip\par\noindent\textbf{A possible culprit: knowledge transfer?}
Nurgalieva and del Rio \cite{NdR} make use of epistemic logic, based on the standard possible-worlds semantics, to analyze the agents' reasoning in the FR paradox, and blame the paradox on what they call the `Trust' postulate. A better name for this principle is `Knowledge Transfer', since it asserts that nested, higher-order knowledge (about another agent's knowledge) allows the transfer of lower-order knowledge between agents. If we read $K_O\varphi$ as ``observer $O$ knows proposition $\varphi$''  (in the sense of ascribing probability $1$ to $\varphi$), then this principle can be formalized as the claim that the following implication is valid (universally true):
$$(\ast) \quad\quad K_O K_{O'} \varphi \Rightarrow K_O \varphi.$$
Nurgalieva and del Rio correctly point out that $(\ast)$ follows trivially from the standard axioms of Epistemic Logic, in particular from
the principles of \emph{Factivity}\footnote{Factivity, also known as `Veracity', is a basic epistemic postulate, that incorporates the main feature of knowledge, and its main difference from mere belief: known propositions are \emph{true}, i.e. in agreement with the facts.} and \emph{Monotonicity}\footnote{In normal modal-epistemic logic, Monotonicity is usually derived from the rule of Necessitation and Kripke's axiom of Distributivity.}:

\smallskip\par\noindent
\begin{itemize}[rightmargin=*]
\item[(\emph{Factivity})]  \,\, \, $K_O \varphi \Rightarrow \varphi$
\item[(\emph{Monotonicity})] \,\, $\mbox{ If } \varphi \Rightarrow \psi \mbox{ is valid, then } K_O \varphi \Rightarrow K_O \psi \mbox{ is valid}$
\end{itemize}

\smallskip\par\noindent
Since they blame the paradox on the Knowledge Transfer principle $(\ast)$, these authors are lead to question the basic assumptions of epistemic logic.

Waaijer and van Neerven in \cite{WvN} similarly reject the idea of `promoting one agent's certainty to another agent's certainty' but add a condition that this rejection applies only when the fact in question `cannot be validated by records from the past'. Lalo\"e \cite{Laloe} refers to a similar idea, suggested by P. Grangier, of finding a way to protect the obtained measurement information from perturbations created by other observers. This can be done by  adding a secret qubit to the experiment, or a `friend of the friend', via which the friend can store the result of his measurement.

The idea of weakening condition (C), by requiring the existence of a persistent record or by letting the information escape outside the lab,
will
form an important ingredient of our own proposed solution below. Yet as we'll see in the next section, this idea needs to be tied in more precisely to the very definition of ``being an observer''.

\medskip\par\noindent\par\noindent\textbf{Counterfactual reasoning about others' observations?}
One should note that one can derive the same contradiction in a way that does \emph{not} make any appeal to the principle of \emph{`knowledge transfer'}. Indeed R. Healey \cite{Healey} proposes such an argument, that does not require any knowledge transfer.
All the inferences (i)-(iii) are now done by $W$ alone (after hearing $\overline{W}$'s announcement).
But note that Healey's version still requires agents to reason about other agents' measurement results (though not necessarily to adopt their perspective, or reason about their knowledge). $W$ reasons by cases, using what Healey calls counterfactual implications of the form :
``If $\overline{F}$ observes `tail' at $t_0$, then the unique
outcome of my measurement of $L$ at $t_3$ would be $fail$.''
On the other hand, $W$ similarly proves that
``If $\overline{F}$ observes `head' at $t_0$, then the unique
outcome of $\overline{W}$'s measurement of $\overline{L}$ would be $fail$, in contradiction to $\overline{W}$'s announcement.''
Together, these lead again to the wrong conclusion that the protocol never halts.

\smallskip\par
The analysis in \cite{Healey} locates the error in the conclusion of step (i), mainly in using the state assigned by $\overline{F}$ to the signal $S$ to derive conclusions about $W$'s measurement of $L$.
According to Healey, ``$\overline{F}$ is justified in using this state assignment for the purpose
of predicting the outcome of a measurement on $S$ only where $S$'s correlations
with other systems (encoded in an entangled state of a supersystem) may be
neglected. But the prior
interaction between $\overline{W}$ and $L$ undercuts $\overline{F}$'s justification for using his state assignment for $S$ to predict anything about $W$'s subsequent measurement of $L$.''

\medskip\par\noindent\par\noindent\textbf{Another suspect: intervention insensitivity?}
Healey thinks the error in this version is the underlying assumption of
``\emph{Intervention Insensitivity}'': \emph{The truth-value of an outcome-counterfactual is
insensitive to the occurrence of a physically isolated intervening event}.
Once again, $\overline{W}$'s measurement of $\overline{L}$ is such a disturbing intervention on the system $\overline{L}$, which \textbf{is} physically isolated from $L$. But this still affects the outcome of $W$'s subsequent measurement of $L$, invalidating step (i) in $W$'s reasoning.

\medskip\par\noindent\textbf{Lack of persistence versus subjectivity: `unstable' facts, or no `facts' at all?}
P. A. Gu\'erin, V. Baumann, F. Del Santo and C. Brukner make an important point in \cite{Brukner2}, according to them we shouldn't simply assume that the (information carried by the) observed outcomes will persist in time. Hence not even $F$ can assume that his `memory of measurement outcomes' from the past and from the present will persist, aka  be still relevant and usable, at all future times. We agree with this point! However, Brukner in \cite{Brukner,Brukner1} gets to even question the existence of objective reality: the ``facts'' (observed outcomes of experiments) are deemed to be ``subjective'' (i.e. observer-dependent).

In contrast, Rovelli \cite{Rovelli,CR} adopts a more realist interpretation, in accordance to which such facts are \emph{relative} (having only a relational meaning, as binary properties relating two systems), but they are nevertheless ``real'' and objective: consciousness, subjectivity, higher-order agency, beliefs etc play no essential role. In \cite{CR}, the authors distinguish between `stable' facts (which \emph{can} be transferred from a reference system to another) and unstable ones. They add that such stability is always only approximative and relative, and that in the case of macroscopic reference systems it can be explained by decoherence.

We tend to agree with the last view, and in fact we think that \textit{there is no need to doubt the ``facts'', but only the probabilistic predictions based on (obsolete such) facts}. A quantum state is just a statistical \emph{summary of the results of past interactions}: there is \textit{nothing subjective} about that. It embodies the relative information that some `observing' subsystem carries about another subsystem. However, the \emph{predictions} derived from such past ``facts'' may be \textit{incorrect}, when they are applied to interactions that have \emph{`erased'}\footnote{In line with standard quantum mechanics, we assume that measurements \emph{can} override or erase quantum information, in the sense that the state-description after the measurement does not contain a copy of the state-description that was true before the measurement. Note that the no-cloning and no-deleting theorems of quantum information theory do not
contradict this assumption, since they only apply to unitary evolutions. In the alternative no-collapse description of such a measuring interaction, what is `destroyed' by the given evolution is the \emph{correlation} between the two subystems: while before this evolution, the two were entangled in such a way that one carried `information' about the other, after the evolution the two have become dis-entangled. If this relative information (carried previously by one system about the other) has not previously been `saved' somewhere else (by another copying entanglement with some safe external `record'), then we can say that this relative information has been `erased' from the universe (-in the sense that it has become completely inaccessible to any future observer). Cf. \cite{Rovelli} for this sense of ``relative information'' carried by a system about another system, and cf. \cite{RovelliNew} for the sense in which such information can be `erased' by a destructive measurement in an orthogonal basis.} {\emph{from the universe all the relative information carried by these quantum states}.
In other words, such past ``facts'', no matter how `objective', \textit{may not necessarily yield `knowledge' about future events, unless (and while) the underlying evidence} (=the quantum information embodied by these facts) \emph{still persists somewhere in the universe}.

\medskip\par\noindent\textbf{Persistence as a criterion of being an `observer'.} Other authors \cite{Jordan} take such stability or persistence of measurement results to be ``a necessary condition on a quantum system to behave as an observer''. This is embodied in \emph{Condition 1} in \cite{Jordan}, which states that: ``For an entity $O$ to behave as an observer, there must exist at least one degree of freedom in which $O$ can encode the results of measurement, and which is untouched for the duration of the considered experiment''.

As stated, this sounds very similar to our proposed solution, and indeed there are close connections between this view and ours. However, the authors of \cite{Jordan} adopt a rather narrow understanding of the stability criterion: namely, they take it to simply forbid even the possibility of wiping the Friend's memory of the measurement results.\footnote{``In particular, to be used in Wigner's friend type scenarios, this condition applies to the friends solely if their memory remain stable even when Wigner makes his own measurements.'' \cite{Jordan}.} This completely trivializes the problems involved in Wigner's Friend and FR paradoxes, and it amounts to imposing a rather brutal `solution': the scenarios involved in these thought experiments simply cannot happen!

\medskip

In contrast, we will adopt a much more liberal interpretation of the persistence criterion: although the initial observer's memory may be wiped out, this doesn't matter as long as records of his observations have already been produced and disseminated throughout the universe; even if each individual such record is wiped out in its turn, this doesn't matter if (the process of copying and leaking is fast enough so that) other records have been produced and disseminated in the meantime; and even if none of this actually happens, it doesn't matter as long as \emph{it may have happened}, as far as another external, background observer can tell.

\section{Our Solution}\label{sec3}

As announced in the last section, our proposal for a solution to Wigner-friend-type paradoxes uses the same idea of \emph{persistence/stability of evidence/information} (as a necessary requirement for sharing it, or for reasoning even counterfactually about such evidence). But in our view this notion is closely linked to the idea of \emph{information leakage}, in the form of continuous production and dissemination of records or copies of this information. Moreover, our solution implements this idea in a `situated' perspective (with respect to a background observer), based on a particular implementation of Relational QM \cite{Rovelli}. The question of informational persistence can thus be weakened to the \emph{mere epistemic possibility of informational leakage}, as far as the observer's knowledge is concerned. It is also reinterpreted, from a metaphysical question about the (in)existence of objective ``facts'', into a pragmatic-epistemic question about the \emph{admissibility of additional observers} (from the perspective of the background observer).

\bigskip\par
We \textit{accept assumptions (S) and (Q)}: the `Cut' between the observer and system is indeed relative in a sense: every system is an ``observer'' from its own perspective, and describes its own interactions with other systems as collapsing measurements of these systems; while from an external perspective, the same interactions can be described as unitary evolutions.
In principle, every `observer' \textit{can} keep its `Cut' \textit{minimal}, by treating every other system as a quantum system (rather than as another collapsing observer).

As already mentioned, we adopt the `situated' perspective of Rovelli's Relational Quantum Mechanics, according to which there is no ``view from nowhere'': quantum state attributions and quantum properties are always relative to a background ``observer'' system $O$. But \emph{in this paper we extend this relational view to the very notions of ``observer'' and ``observation''}; moreover, to make them more realistic (i.e. applicable to real-life situations), these notions have to be \emph{further relativized to a given history or set of histories}, that encompass the relevant timeframe and the available actions.\footnote{In this sense, our approach borrows some of the features of the ``consistent histories'' interpretation.}

\medskip\par
This means that \textit{assumption (C) will have to be weakened}: sharing information between different `observers' (or reasoning counterfactually about other systems as `observers') is only justified when there exists a valid non-minimal `cut' that includes all of these subsystems on the ``observer'' side (together with the reasoner). For this, we need \emph{a criterion for deciding whether a given system can be regarded as an ``admissible observer'' from the perspective of a background observer $O$ and with respect to a given history or set of histories}.

The main idea behind our solution to the paradoxes is that the needed criterion has to do with the \emph{persistence of the information} possessed by the subsystem in question. Roughly speaking, a subsystem $A$ can count as an admissible observer for (another subsystem) $O$ only if,  as far as $O$ ``knows'', all the information ever possessed by $A$ (at any moment of the relevant set of histories) may ``forever'' survive \emph{somewhere} in the universe.\footnote{``Forever'' needs not be understood in an absolute sense: it is relative to the relevant timeframe, history or set of histories.} This doesn't mean that $A$'s own memory is necessarily perfect, or that it is immune from external erasure: it just means that, even if $A$'s memory has been wiped put, $O$ can still never be sure that $A$'s information has not survived somewhere else (e.g. in the form of a `leaked copy'', or a copy of a copy etc).

\smallskip

In the next subsection we sketch a formal rendering of our solution.

\subsection{Formalization}\label{subsec31}

First, we are given a (finite or infinite) list of labels (letters) $\Sigma=\{s_1, s_2, \ldots\}$, denoting \textit{elementary systems}, together with associated Hilbert spaces $H^{(s)}, \dots$. From these, we can generate composite \emph{systems}, given by subsets $S, O, S', A, B, \ldots\subseteq \Sigma$, each with its associated Hilbert space $H^{(S)}:= \bigotimes_{s\in S} H^{(s)}$.
We leave unspecified \emph{which} subsets $S\subseteq \Sigma$ are admitted as ``systems'', in order to accommodate various approaches: according to one approach \emph{all} subsets are systems (and hence in particular the universe $\Sigma$ is itself a system), while according to others some restrictions (e.g. finiteness) are to be imposed in order for a subset $S\subseteq \Sigma$ to qualify as a system.

\smallskip

The \emph{state} of any system $S\subseteq \Sigma$ from the perspective of some `observing' system $O\subseteq \Sigma$ is denoted by $S_O$. When we want to make explicit the time of this state attribution, we write $S_O^t$ (for the state of $S$ at time $t$ from the perspective of $O$), where we assume given a discrete set of temporal moments $t=0, 1, \ldots$.  All systems $S$ (even including the ``supersystem'' $\Sigma$) are always described from such a `situated' point of view, associated to a background observer $O$. The described system $S$ may or may not overlap with the `observing' system $O$, or we may even have $O\subseteq S$. In case they do overlap, the ``observed subsystem (of $S$)'' is the non-overlapping part $S-O$ of the system under description.

This means that we allow the `observer' $O$ to describe/represent systems $S$ that it may not observe (since they may include parts $S\cap O$ that overlap with itself). While it is commonly assumed that no system can observe itself, there is no reason to deny an observer the capacity to `know' or represent its own state. Indeed, if we identify this state with the observer's \emph{memory} (e.g. the record of all $O$'s measurement results), then having this capacity amounts to having ``perfect recall'' (a standard assumption in epistemic logic). This may be considered as a simplifying idealization applying only to `rational' agents, but it is a useful one. In quantum-mechanical terms, this amounts to assuming the following Introspection postulate (I):

\bigskip\par\noindent
\begin{itemize}[leftmargin=*]
\item[(I)] the state $O_O^t$ (assigned to $O$ itself from $O$'s own perspective at any given time $t$) is always a \emph{pure, fully separated state} of $H^{(O)}$, i.e. one of the form $\bigotimes_{i\in O} s_i^t$, where each $s_i^t$ is a pure state of $S^{(i)}$ (for every $i\in O$). In fact, we typically assume that $s_i^t$ are elements of the standard basis for $H^{(i)}$ (though this is of course just a matter of convention): intuitively, $O$'s `memory' consists of a number of `bits' $s_i$, recording the results of past measurements.
\end{itemize}

\medskip

The principle $(Q)$ of universality of Quantum Mechanics amounts in this setting to assuming the following postulates:

\bigskip\par\noindent
\begin{itemize}[leftmargin=*]
\item[($Q_D$)] at any moment $t$, the state $S_O^t$ of any system $S$ according to any `observer' $O$ is given by a density operator $\rho_S^t$, representing a (pure or mixed) state $S_O^t$ over $H^{(S)}$;
    \item[($Q_+$)] for every subsystem $A\subseteq S$, we can compute its state with respect to $O$ at any time $t$ from the corresponding state $S_O^t$ of the super-system $S$ (by using the fact that $S$ can be written as a composed system $S= A+ (S-A)$, and thus by QM we have): $A_O^t = Tr_{\,S-A} (S_O^t)$, where $Tr_{\,S-A}$ is the partial trace operator wrt subsystem $S-A$;
\item[($Q_U$)] for any moment $t$ at which the observer $O$ does \emph{not} interact with any part of a system $S$, the evolution of $S$ at time $t$ is given by $S_O^{t+1}= U^t (S_O^t):=U^t \rho^t (U^t)^\dagger$, where $U^t$ is some unitary map on $H^{(S)}$ (and $\rho^{t}$ is the density operator representing the state $S_O^t$ of $S$ at time $t$);
\item[($Q_P$)] for any moment $t$ at which the observer $O$ \emph{does} interact with some external subsystem $A\subseteq S-O$, the evolution of $S$ at time $t$ is given by $S_O^{t+1}= (P^{t}_{\rho_A}\otimes  B^{\rho_A}_O \otimes I_{S- (O\cup A)})(S_O^t)$, where $I_{S-(O\cup A)}$ is the identity operator on $H^{(S-(O\cup A))}$, $P^t_{\rho_A}$ is some projector\footnote{Projectors are self-adjoint, idempotent linear maps, that are used within the Hilbert space formalism to represent the physical process called `projective measurements', as described in Quantum Mechanics, see e.g. \cite{Nielsen}.} onto some pure state $\rho_A:=\rho_A^t$ of the system $A$, and $B^{\rho_A}_O$ is the operation of augmenting $O$'s memory $O_O^t$ with (some bit-encoding $B^{\rho_A}$ of) the outcome $\rho_A$ of the measurement of $A$ by $O$.
\end{itemize}
\bigskip\par\noindent

The combination of the postulates ($I$) and ($Q_+$) implies that the state $\Sigma_O^t$ of the universe according to $O$ (if such a state exists) must be at any given time $t$ a separable state of the form $(\Sigma-O)_O\otimes O_O$, where $O_O=\bigotimes_{i\in O} s_i$ is the above-mentioned fully separable state. More generally, this poses certain constraints on the relative state of every subsystem $S\subseteq \Sigma$, namely the state $S_O^t$ must at every time be of the form $(S-O)_O\otimes (S\cap O)_O$ (where $(S\cap O)_O$ is the restriction $\bigotimes_{i\in S\cap O} s_i$ to $S\cap O$ of the above-mentioned fully separable state $O_O$).

\medskip

These assumptions lead us to the following framework.

\smallskip\par\noindent\textbf{Histories and protocols.} A \emph{history (of length $n$) of a system $S$ wrt an observer $O$} is a sequence $h=(S_O^0, T^1, S_O^1, T^2, \ldots, S_O^{n-1}, T^n, S_O^n)$, where for all times $k<n$, $S_O^k$ are state descriptions of the system $S$ (with respect to $O$) s.t. $S_O^{k} =T^k(S_O^{k-1})$, and where all $T^k$ are linear maps (called \emph{dynamical maps}), required to satisfy the postulates ($Q_U$) and ($Q_P$) above; i.e. $T^k$ are either unitary maps on $H^{(S)}$ or they are `projective' maps (of the form $P_{\rho_A}\otimes  B^{\rho_A}_O \otimes I_{S- (O\cup A)}$, modelling an $O$-measurement of some subsystem $A\subseteq S-O$ that collapses the observed system $A$ onto a pure state $\rho_A$, while recording it as $B^{\rho_A}_O$ into $O$'s memory).

Given any time moment $t\leq n$, the \emph{$t$-restriction of the history} $h=(S_O^0, T^1, \ldots, S_O^n)$ is the initial $t$-long subsequence $h|t:= (S_O^0, T^1, \ldots, T^t, S_O^t)$.

 A \emph{protocol} (for $S$ wrt an observer $O$) is a family $\pi$ of histories of the same length $n$ and having the same initial state $S_O^0$.

\medskip

In principle, the ``Cut'' between observer and observed systems is arbitrary: any system can play the role of observer $O$ \emph{from its own ($O$'s) perspective}. But the interesting question is: \emph{which other systems can be admitted as ``observers'' from $O$'s perspective}? The answer is far from obvious, and formulating it is the core contribution of this paper.

\medskip

\medskip\par\noindent\textbf{Propositions about evolving system.} In the following, we'll consider \emph{properties of an evolving system} $S$, always considered from the perspective of some observer $O$.
Formally, given a protocol $\pi$ (of length $n$) for $S$ wrt $O$, such properties will be captured propositions $\varphi$ that may hold at
some moment $t\leq n$ of some history $h\in \pi$; we write $(h,t)\models_\pi \varphi$ in this case.

Evidently, the set of propositions will be closed under the standard Boolean operators. But we can also introduce \emph{epistemic} and \emph{temporal} operators. We only give here the formal definition of our epistemic operator (though temporal notions will be implicitly used in our definitions of persistent records informational persistence below):

\medskip\par\noindent\textbf{Propositional Knowledge.} Given a protocol $\pi$ for $S$ wrt $O$, a history $h\in\pi$ (of length $n$) and a time moment $t\leq n$ , we say that \emph{a proposition $\varphi$ is known (by $O$) at time $t$ of history $h$}, and write $(h,t)\models_\pi K_O\varphi$, iff $\varphi$ is true at time $t$ of all histories $h'$ that agree with $h$ on their initial $t$-sequence. Formally:
$$(h,t)\models_\pi K_O\varphi \,\, \mbox{ iff } \,\, (h', t)\models \varphi \mbox{ for all $h'\in \pi$ s.t. $h|t= h'|t$}.$$
The possibilistic dual of knowledge $\hat{K}_O$ is defined as usual in epistemic logic by putting
$$\hat{K}_O \varphi \,\, := \,\, \neg K_O \neg \varphi,$$
where $\neg \varphi$ is the (classical) negation of $\varphi$. It is easy to see that $(h,t)\models_\pi \hat{K}_O\varphi$ iff there \emph{exists} some history $h'\in \pi$ s.t. $h|t=h'|t$ and $(h',t)\models\varphi$. In English, we can read $\hat{K}_O \varphi$ as saying that \emph{$\varphi$ may hold as far as $O$ knows} (at time $t$ of history $h$), or that $O$ considers $\varphi$ \emph{possible} (at time $t$ of $h$).

This definition of `knowledge' essentially incorporates the notion that \emph{all that our observer $O$ `knows'} about a system $S$ at a given moment $t$ follows from two sources: (1) \emph{her own records of past interactions} with $S$ up until the current moment $t$, as embodied by the initial $t$-subsequence of the current history (-i.e., the sequence of states of $S$ wrt $O$ and the dynamical transformations up until $t$); and (2) \emph{the protocol itself} ($\mbox{-giving}$ background information about the possible histories, possibly including some information about the future course of these histories after $t$).  Indeed, any information that $O$ might have about $S$ that is not gained via past interactions during the protocol (-e.g. based on global constraints, or on theoretical, \emph{a priori} grounds) can be included in the description of the protocol itself (since it amounts to constraints on the set of possible histories of $S$).

\medskip\par\noindent\textbf{Examples.} An example of a property that is known by $O$ at time $t$ of a history $h=(S_O^0, T^1, S_O^1, T^2, \ldots, S_O^n)$ is (the proposition expressing) \emph{the (mixed) state} $S_O^t$ of the system $S$ at time $t$. Moreover, the above definition makes it clear that $O$ \emph{knows all the past history} of $S$ up until moment $t$: this is natural, since this history represents $O$'s own description of the past evolution of $S$, recording the results of all past measurements. It is unsurprising to assert that this initial part of the history is known to $O$ at $t$: after all, the density operators and dynamic matrices forming this sequence are nothing but \emph{$O$'s own description of what has happened to $S$}, \emph{from her own perspective}. While this notion of knowledge is qualitative, it can also encode \emph{probabilistic knowledge} as well. For instance, if the density operator $\rho=S_O^t$ represents a pure state, then the observer can use Born's rule to compute the probability of any particular outcome of any measurement that might be performed on $S$ at this time: $O$ ``knows'' the probabilities of these outcomes.

\medskip\par\noindent\textbf{Proviso: no nested knowledge yet!} The above definition does not entitle us yet to assign a meaning to formulas of the form $K_O K_A \varphi$, where $A$ is any other subsystem of $S$ (different from the observer $O$). For such a nested formula to make sense, we will first need to determine \emph{whether or not the system $A$ may count as an `observer' from $O$'s perspective}: formulas of the form $K_A\varphi$ will belong to the observer $O$'s inner language (and thus be allowed in the scope of an epistemic $K_O$-operator) only if $A$ is admissible as an observer according to $O$.

\medskip\par\noindent\textbf{Records.} Given systems $A, B\subseteq S$ of the same dimension, we say that \emph{$B$ is a record of $A$} in state $S_O^t$ (of system $S$ at time $t$ according to $O$) if there exists some unitary map $U_{A\to B}: H^{(A)}\to H^{(B)}$ s.t. for every element $e_A$ of the (standard) basis of $H^{(A)}$
we have $Tr_{S-B}( (P_{e_A}\otimes I_{S-A})(S_O))= U_{A\to B}(e_A)$ (where $P_{e_A}$ is the projector onto $e_A$, $I_{S-A}$ is the identity operator on $H^{(S-A)}$, and $Tr_{S-B}$ is the partial trace wrt the complement of $B$). Intuitively, whenever system $A$ stores `relative information' (results of past measurements) in the form of the classical bits forming $e_A$, the same information is encoded in system $B$ (via the reversible encoding $U_{A\to B}$).\footnote{Note that, although the above definition refers to the standard basis for $H^{(A)}$, the concept itself is independent of the choice of basis, because of our existential quantification over unitaries. A change of basis will only replace the unitary $U_{A\to B}$ with some other unitary $U'_{A\to B}$.} Here, systems $A$ and $B$ may be disjoint or may overlap, and in the second case they may even be identical: indeed, in any state $S_O$, \emph{every subsystem $A\subseteq S$ can be thought of as a `record' of itself}.

\medskip\par\noindent\textbf{Persistent records.}
Given a protocol $\pi$ for a system $S$ wrt $O$, we say that a subsystem $A\subseteq S$ \emph{has a persistent record at time $t$} if there exists a subsystem $A'\subseteq S$ (of the same dimension as $A$) s.t. (1) $A'$ is a record of $A$ in state $S_O^{t-1}$, and (2) if $T_S^t$ is the unique dynamical map associated to $h$ at time $t$ (or any of the dynamical maps associated to any history $h\in \pi$ at time $t$), $E$ is any basis for $H^{(A')}$ and $F$ is any basis for $H^{(S-A')}$, then there exist some unitary map $U_{A'}^t$ and some dynamical map $T_{S-A'}^t$, s.t. $T_S^t(e_{A'}\otimes f_{S-A'})= U_{A'}^t (e_{A'}) \otimes T_{S-A'}^t (f_{S-A'})$ for all basis elements $e_{A'}\in E, f_{S-A'}\in F$.

\medskip\par\noindent\textbf{Informational persistence throughout a history/protocol.} A subsystem $A\subseteq S$ \emph{is (informationally) persistent throughout a protocol} (for the system $S$ wrt $O$) if we have that: at every time $t\geq 1$, $A$ has some persistent record $A'_t$;  each such record $A'_t$ has itself a persistent record $A''_t$ at time $t+1$; and so on (for all times $t+k\leq n$, where $n$ is the length of the given history or protocol).

\medskip\par\noindent\textbf{Admissible observers.} A subsystem $A\subseteq S$ is an \emph{admissible observer wrt (a background observer system) $O$ and a given history/protocol} (for $S$ wrt $O$) iff, for all that $O$ knows (at any given time), $A$ \emph{may be informationally persistent} throughout the history/protocol; in other words, if $O$ never gets to know for a fact that $A$ is not (will not be, or has not been) persistent throughout the history/protocol.}
\medskip

The main idea of our solution, is that, only \emph{after} this requirement (of $A$ being an ``admissible observer'')  is checked (using $O$'s Minimal Cut), $O$ becomes entitled to extend its Cut, to include $A$ on the `observer' side.
In other words, a system $A$ can be admitted as an `observer' (from $O$'s perspective) only if it can never be known for sure (by $O$) that the information once carried by $A$ has been (or will be) completely erased from the universe (at least not for the duration of the given history/protocol).

Note that the notion of ``admissible observer'' is always relative to a background observer system $O$ and to a given history/protocol. Also note that the informational persistence of $A$ does \emph{not} require that past states of $A$ are preserved \emph{within} $A$. It is perfectly compatible with the possibility that past information is erased at some point from $A$'s `memory' (or even that its whole memory is `wiped out', by being reset to some default state), as long as that past information is still saved \emph{somewhere} in the universe (in possibly encoded form via some unitary transformation).

\medskip

Furthermore, one should note that actual informational persistence is \emph{not} necessary for being an admissible observer. All that is needed is that the background observer $O$ \emph{cannot exclude} the possibility of informational persistence: it \emph{cannot ensure} (with probability $1$) that the relative information carried by the admissible observer is completely erased from the universe (at least not for the duration of the relevant history/protocol). This is important: actual persistence (of all the information ever acquired by a subsystem $A$) is a strong condition, conferring some kind of ``immunity guarantee'' or indestructibility to $A$'s information, which seems to demand that $A$ has a high degree of power or control over information, or at least that $A$ has an inexhaustible ability to produce records and spread them far away from any relevant interference from $O$. In contrast, the \emph{mere epistemic possibility} of $A$'s informational persistence (according to $O$'s state of knowledge) is a much weaker condition, which has more to do with the \emph{limits of $O$'s power of controlling information}: it only means that $O$ cannot prevent $A$'s information from ``leaking'' beyond $O$'s control.

\medskip

Note that by the above definition, $O$ is an admissible observer wrt $O$ itself and any given history/protocol for $S+O$ (wrt $O$). We abbreviate this by saying that \emph{any system $O$ is always an admissible observer wrt itself}. In \emph{this sense}, the `cut' between the observer and observed is indeed arbitrary!

\bigskip\par\noindent
\begin{proposition}

Given a protocol $\pi$ for a system $S$ wrt an observer $O$, let $A\subseteq S$ be an admissible observer. For every description of the state evolution of the system $S$ wrt $O$ (as defined above), there exists an alternative description of the evolution of $S$ wrt $A$ (in which $A$ takes the place of the background observer), such that the two descriptions are `compatible' in a weak sense, namely relative to the protocol $\pi$.
\smallskip\par\noindent
Here, the notion ``compatibility'' of state descriptions is weaker than compatibility in the sense of Leifer and Spekkens \cite{LF}, and it is only {\emph{relative to the given protocol}: at each time $t$, the two state descriptions can be refined to states that \emph{assign the same probabilities to all possible outcomes of all the measurements that can happen according to the protocol $\pi$}.}
\end{proposition}
\medskip

The \emph{proof} of this result is essentially based on an elaboration of the argument given in our analysis of the `leaking' scenario 2 in Section \ref{subsec22}. Spelling out the proof in full generality requires further formalities, that go beyond the scope and context of this paper, and are best left for a future technical publication in a Formal Logic journal.

\medskip\par\noindent\textbf{Nested knowledge.} Intuitively, $O$ can reason about $A$ \emph{as an observer} only if $A$ is an admissible observer wrt $O$: only in this case, a nested formula of the form $K_O K_A\varphi$ is meaningful (provided that $\varphi$ does not contain any inner epistemic operators). Further nesting requires further conditions: for a formula of the form $K_O K_A K_O \varphi$ to be meaningful, each of the two systems $A$ and $O$ has to be an admissible observer for the other system. We now generalize this notion to arbitrary groups.

\medskip\par\noindent\textbf{Communities of observers.} A group $\mathcal{G}$ of subsystems of a system $S$ (that includes $O$ itself) is a \emph{community of admissible observers} wrt $O$ (and a given history/protocol) if every system $A\in \mathcal{G}$ is an admissible observer both wrt to any other system $B\in \mathcal{G}$ as well as wrt $O$ (and the given history/protocol).

Intuitively, a group of `agents' forms a community of observers if there is a kind of ``fair balance of powers'' between them: each agent's ability to knowingly control and erase information from the universe (or to at least know for a fact that it has been fully erased) cannot fully overwhelm other agents' ability to produce and disseminate records of their own information throughout the universe.

\medskip

The above Proposition has the following immediate consequence:
\bigskip\par\noindent
\begin{corollary}
Given a protocol for a system $S$ wrt an observer $O\subseteq S$, let $\mathcal{G}$ be a community of observers s.t. $O\in \mathcal{G}$. Then,
for every description of the state evolution of the system $S$ wrt $O$ (as defined above), there exist compatible descriptions of the evolution of $S$ wrt to each of the admissible observers $A\in \mathcal{G}$, and moreover all these descriptions are mutually compatible.
\end{corollary}
\bigskip\par\noindent

\medskip\par\noindent\textbf{Indefinite nesting: the conditions for full epistemic logic, knowledge transfer and quantum computation protocols.} Intuitively, whenever we have a community of observers $\mathcal{G}$, then each of them can reason about the others' observations as well as about the others' reasoning about itself and all others, etc. In other words, \emph{only for such a community of observers we can meaningfully use all the power of epistemic logic}, with indefinitely long nested levels of knowledge, e.g. $K_O K_A K_B K_A K_C \varphi$, etc. All the standard axioms of Epistemic Logic hold in this case, and so in particular the principle of Knowledge Transfer holds (since it is a theorem of Epistemic Logic).
As a consequence, the agents involved in any standard protocol studied in the fields of Quantum Information and Quantum Computation need to always be assumed to form a community of observers.

\medskip

One should stress that, from the perspective of an external observer $O$, checking whether or not another system $A$ satisfies the conditions for being an ``admissible observer'' should be done \emph{without} first assuming that $A$ collapses the state of the systems it interacts with! In other words, $O$ is entitled to reason ``from $A$'s perspective'' only \emph{after} the conditions of being an ``admissible observer'' are verified. One cannot `prove' these conditions by first assuming $A$'s collapsing perspective, and then using this assumption to show that $A$ satisfies the conditions for being an admissible observer: that kind of circular reasoning would be a case of \emph{petitio principii}! As we'll see, this is exactly the kind of reasoning that is involved in the usual descriptions of the FR Paradox.

\subsection{Application to Wigner's Friend-type scenarios}\label{subsec32}

According to the above analysis, whether or not a system is an admissible observer depends on some background observer \emph{and} on the background protocol. One of the problems with Wigner's Friend-type of scenarios is that this background information is usually not fully specified.

\bigskip\par\noindent\textbf{Scenario 1, revisited.}
First, consider the original Wigner's Friend paradox (scenario 1). Who should we take as our background reference `observer'? In the first case, it seems natural to take it to be $W$, while in the second case it could be either $W$ or $O$. This is OK \emph{only} as long as the scenario does \emph{not} involve any wiping out of the information possessed by systems $W$ or $O$. Let us grant that. Next, to decide whether or not $F$ is himself an ``admissible observer'' from the perspective of $W$ (or $O$), we need to know \emph{what happens after that} (-what is the subsequent history, or the set of possible histories), or at least we need to know \emph{what do any of these background observers know about what will happen}. Essentially, Wigner Friend's story is unfinished: it all depends on what may happen subsequently! We can consider a number of cases, but in each case, once the protocol is fully specified, then there is a unique conclusion.

\medskip\par\noindent\textbf{Scenario 1, Case 1.} Suppose that, in scenario 1, $W$ \emph{knows that no information will be leaking} out of the lab $L$ and that moreover \emph{it is possible for himself ($W$) or for some other external agent to perform a `destructive' measurement of the lab $L$} as a whole (e.g. a measurement in a basis that contains $|fail\rangle_L$), that undoes the correlation between $S$ and $F$. This means that, at some later moment during the protocol, $W$ may come to know that the information carried by $F$ about $S$ (via the given correlation) has been completely ``erased'' from the universe (in the sense it is made completely inaccessible to any possible observer). In this case, $F$ simply does \emph{not} meet the conditions for being an admissible observer wrt $W$: after such a destructive interaction, $W$ will know that no record of $F$'s information persists till the end of the protocol. No contradiction ensues: according to our analysis, $W$ is \emph{not} entitled to consider $F$'s interactions as collapses of the state space by some admissible observer. From $W$'s perspective $F$ is just a quantum system, whose interactions are governed by unitary evolutions.

\medskip\par\noindent\textbf{Scenario 1, Case 2.} As in case 1, $W$ \emph{knows} that no information will be leaking out of the lab $L$, but now
the protocol is set up such that $W$ \emph{can never get to know that any `destructive' interaction of the above kind has occurred}. This can be because e.g. the protocol does not allow for a global measurement of $L$ as the one above (or because the lab $L$ is too large to be subject to such a precise measurement, or more generally because $W$ does not have full quantum control over the lab $L$ due to the interference of other factors, and so $W$ cannot ensure that the destructive measurement succeeds). In that case, for all that $W$ knows, $F$'s information about $S$ may persist till the end of the protocol: $F$ plays the role of being \emph{its own persistent record} (of his own information about $S$), and thus he \emph{is an admissible observer wrt $W$}. So $W$ is then able to reason about $F$'s interactions as collapsing measurements, and the predictions given by this description \emph{are} compatible with the predictions based on $W$'s minimal cut, as far as the (probabilistic prediction of every outcome allowed by the) background protocol is concerned.

\bigskip\par\noindent\textbf{Scenario 2, revisited.} This brings us to scenario 2, in which $F$'s measurement results \emph{are leaked} in the environment (e.g. to the third `observer''), \emph{before} any global measurement of the lab $L$ could take place. Even if we take $W$ to be our background observer, and even if $W$ \emph{doesn't know} that $F$'s information has leaked, he still must \emph{consider the possibility} that this may have happened: assuming the opposite in this context would be an example of relying on a false, ungranted assumption, and thus an example of faulty reasoning! If the leakage/communication does happen, then $W$ \emph{cannot know} that it didn't happen; and even if the leakage didn't happen, as long as $W$ cannot be sure of this, he has to take into consideration this very possibility. And \emph{this very possibility establishes one of the necessary conditions for $F$ to be an admissible observer wrt $W$} in scenario 2.

But this is just a necessary condition: whether or not $F$ really is an admissible observer \emph{depends again on everything that may happen next}, in all the possible histories permitted by the protocol under consideration. So we have to look once again at the possible full specifications of the protocol, which can be grouped in a number of cases.

\medskip\par\noindent\textbf{Scenario 2, Case 1.}
Suppose for instance that $W$ knows that it is possible to perform next a global measurement of the super-system $L+O$ \emph{before} any information leaks out of this super-system $L+O$. In that case, $F$ is \emph{not} an admissible observer wrt $W$. In fact, this scenario already contradicts the implicit assumption that the external system $O$ was itself an admissible observer (wrt $W$): neither $F$ nor $O$ are admissible observers in this case!\footnote{So, as a side note, we should stress that using the letter $O$ for that external system in Scenario 2 does \emph{not} automatically make it an admissible observer (-in the same way in which denoting the internal subsystem by $F$ and calling it a ``Friend'' did not automatically make it an admissible observer). Whether $F$ or $O$ can be really be considered observers from $W$'s perspective
is an issue that has to be first settled by looking at a full specification of the relevant protocol. Before that, $W$ is \emph{not} entitled to considered their interactions as collapsing measurements.}

\medskip\par\noindent\textbf{Scenario 2, Case 2.}  In contrast, suppose now that the protocol was set up such that $W$ \emph{can never get to know that any such `destructive' interaction of the super-system $L+O$ (that would `erase' $O$'s information about $F$) has ever occurred}. So, for all that $W$ knows (at all moments), $O$ \emph{may provide a persistent record} of $F$'s information. In that case, $F$ \emph{is an admissible observer} wrt $W$, and once again $W$ can reason about $F$'s interactions as collapsing measurements, and the predictions given by this description \emph{are} compatible (wrt the given protocol) with the predictions based on $W$'s minimal cut.

\bigskip\par\noindent\textbf{An analysis of the FR Paradox.} Unlike Wigner's Friend scenario, the FR Paradox comes with a full specification of the underlying protocol. We can safely assume that the external observers $W$ and $\overline{W}$ form a community of mutually admissible observers, since the scenario does not mention any measurements of super-systems that encapsulate them. But what about $F$ and $\overline{F}$?
At first sight, it may seem that $\overline{F}$'s measurement results are `leaked' out of $\overline{L}$ via the signal $S$; moreover, this seems to happen ``just in time'' i.e. before $\overline{W}$ performs his global measurement of $\overline{L}$.

But this is not quite true. The signal $S$ does not satisfy our conditions for being a ``persistent record'' of $\overline{F}$'s information, indeed we will argue that it is not a record at all! Indeed, the usual story \emph{already assumes} that $W$ or $\overline{W}$ can reason from $\overline{F}$'s and $F$'s perspective
\emph{as collapsing observers}, then argues that they can first infer that $F$ measured `up', and then conclude that $\overline{F}$ must have measured `tails'; and only based on those inferences, one could then argue that the information carried by $\overline{F}$'s result (`tails') has been indeed leaked to $F$, and made available to the external observers. But this is exactly the kind of \emph{petitio principii} mentioned above:
according to our solution, $W$ and $\overline{W}$ are \emph{not} entitled to reason by cases about $F$'s and $\overline{F}$'s measurement results \emph{until} we establish that $F$ and $\overline{F}$ are admissible observers!

But to do that, we have to first reason from the perspective of $W$'s and $\overline{W}$'s (joint) ``minimal cut'', according to which they themselves are the only collapsing observers, while $F$ and $\overline{F}$ are just quantum systems (whose interactions within the labs and with each other are `no-collapse', i.e. unitary evolutions). If we do this, we can easily see that the signal $S$ does \emph{not} formally constitute a record of $\overline{F}$'s information: the entanglement between $L$ and $S$ does not carry enough information to recover the correlation between $R$ and $\overline{F}$. This is in fact reflected at an informal level in the following intuitive observation: even if we imagine that $\overline{F}$ and $F$ would perform collapsing measurements, one can recover $\overline{F}$'s hypothetical measurement result \emph{only in one history branch }(corresponding to one of the hypothetical measurement results by $F$, namely `up', which is indeed compatible only with `tails'); while in the other branch, an $F$-result of `down' is compatible with \emph{any} measurement result by $\overline{F}$, and thus \emph{it does not carry any definite information} about $\overline{F}$'s state.

This simple, intuitive observation explains in a sense why, when we adopt a non-collapse view about the interactions performed by $F$ and $\overline{F}$ (in effect putting the above-mentioned branches back together into an entangled whole), the resulting unitary evolution does \emph{not} make $S$ into a true record of $\overline{F}$'s information. Though we skip the details here, this fact can be formally verified by performing the unitary calculations and noting that in the resulting entangled state of $L+\overline{L}$, $S$ does \emph{not} satisfy our definition for being a `persistent record' of $F$ at this time. Thus, the actions of sending the signal $S$ to the other lab $L$ and of $F$ ``measuring'' the signal at time $t_1$ do not constitute a true ``leak'' in our sense, since the entanglement they establish between the two labs does not endow $L$ (and $F$) with a record of the correlation between $R$ and $\overline{F}$: there simply is not enough information in $L$ at this stage to fully recover that correlation.

Since at the next stage (at $t_2$) a global measurement of $\overline{L}$ is performed by $\overline{W}$ (and moreover this measurement is a destructive one, since it is done in a orthogonal basis, thus wiping out $\overline{F}$'s memory), and since this happens \emph{before} $\overline{F}$'s information has been fully copied/recorded in any other safe place, we can only conclude that\emph{ $\overline{F}$ fails to satisfy the conditions for being an admissible observer} wrt $W$ or $\overline{W}$.

The same applies even more straightforwardly to $F$: $F$'s information is not leaked anywhere outside of $L$ before $W$ performs a global destructive measurement of $L$. So \emph{neither $F$ nor $\overline{F}$ are admissible observers} for $W$ and $\overline{W}$.

Since the four systems do \emph{not} form a community of admissible observers, we cannot apply the full power of Epistemic Logic to all four ``agents''. We cannot use nested modalities of the form $K_W K_F \varphi$ etc, to allow the external observers $W$ and $\overline{W}$ to reason about the internal systems $F$ and $\overline{F}$ \emph{as observers}. The only community of observers that we have in this scenario that contains more than one observer is $\{W, \overline{W}\}$. So we can only allow nested operators of the form $K_W K_{\overline{W}} K_W \ldots$: $W$ and $\overline{W}$ can indeed reason about each other as observers, but they should only reason about $F$ and $\overline{F}$ as non-observing quantum systems.

\medskip

\par\noindent\textbf{Alternative Scenario: FR with leaking.} In contrast, suppose that we change the scenario of the FR Paradox, by ensuring that $F$ and $\overline{F}$ are immediately `leaking' their information (by sending a record into some remote environment, over which $W$ and $\overline{W}$ have no full control), \emph{before} any global measurements can be performed on $L$ or $\overline{L}$. In this case, according to our definitions, both $F$ are $\overline{F}$ \emph{become admissible observers} (wrt $W$ and $\overline{W}$). And indeed, it is easy that, in this modified scenario, the paradox disappears!

\section{Conclusions}\label{sec4}

Our approach in this paper is based on a clarification of the type of systems that can be viewed as a true agent and can act as an `observer'.
At a first approximation, \emph{only leaking systems (whose information is never completely erased from the universe) are ``true agents'' (to whom we can attribute ``knowledge'', and reason about it using epistemic logic)}. This doesn't mean that their memory is perfect, or that it is immune from external erasure: it just means that their information \emph{survives somewhere ``forever''} (e.g. by being copied in other locations, before being erased from any given location).
To put it conversely: \emph{a system can be treated as an observer only for a period in which its information is preserved somewhere (throughout the period)}.

But this ``absolute'' notion of observer is just a first approximation, and as such it is too strong. A deeper, more sophisticated rendering of our solution makes the notion of ``observer'' to be itself \emph{relative} to some other, background observer $O$. Moreover, the strong, absolute requirement of informational persistence is replaced with a weaker, \emph{epistemic }condition, namely that the background observer $O$ cannot exclude the possibility of informational persistence: she simply is not in a position to know for certain that the relevant information is (has been or will be) fully erased from the universe.

Finally, we have the third, fully correct, approximation of our solution, which makes the notion of ``admissible observer'' to also be relative to a given history or set of histories (a `protocol'). Thus, ``forever'' is not understood here in an absolute sense: it is relative to a given period of time, a history, or a given protocol. A system $A$ is an admissible observer for $O$ as long as the background observer $O$ cannot exclude the possibility that $A$'s information may survive throughout the protocol.

When we consider a \emph{family} of subsystems, and want to be able to allow them to use nested epistemic operators and the full power of Epistemic Logic to reason about each other as observers,
we need to first ensure that each of them is an ``admissible observer'' with respect to all the others. In our technical terms, they need to form a ``community of admissible observers''. Intuitively, this means that none of them is able to `control' (i.e. fully erase from the universe, or at least know for sure that it's been fully erased) the information of any of the
others.

\medskip

This context-dependent and observer-dependent approach gives us a pragmatic-epistemic solution to Wigner Friend's-type paradoxes: there are no absolute, universally acceptable observers, but only (communities of) admissible observers relative to some background observer and background timeframe or protocol. Every system is an admissible observer to itself, but not every system is an admissible observer to every other system. And it is quite possible that no subsystem is an admissible observer with respect to the whole (past and future) history of the universe: if for instance all current local correlations between subsystems will eventually be completely wiped out in the far future, then no current ``agent'' will be an admissible observer with respect to any long enough history. But, in the context of a restricted history or set of histories available and relevant to a given background observer, there will typically exist other admissible observers.

\medskip

Concluding from this that Epistemic Logic is simply inconsistent with Quantum Mechanics (as some authors seem to suggest) sounds rather far-fetched and unhelpful: after all, many of the multi-agent protocols in Quantum Information and Quantum Computation make implicit use of epistemic reasoning about other observers.
Our conclusion is rather that Epistemic Logic is applicable only to groups of physical systems that satisfy certain conditions (namely, they form a community of admissible observers), and that moreover these conditions are typically satisfied only in a relative sense (from the perspective of a background observer, and for the duration of a relevant history or protocol).

\smallskip\par
In other words, sharing observations between systems, or reasoning counterfactually about other systems' observations, is not always possible or justified, but
\textit{only when there is a meaningful possibility that these subsystems are ``leaking'' (accidentally or systematically)}: for all we know, their information may be continuously copied and disseminated somewhere else (by entanglements with a large part of their environment), so that we cannot completely erase this information from the universe.

\smallskip\par
The more ``leaky'' or open a system is, the harder it is for us to encapsulate it and control its information, the more likely it is to be an admissible observer for us.
This is our explanation for the fact that standard epistemic
reasoning \emph{can} typically be applied without any paradox in common macroscopic situations (such as
multi-agent quantum protocols): all macroscopic observers are typically open,
``leaking'' systems.

\bigskip\par
To conclude, the view adopted in this paper is that, on the one hand, every system can be treated as an `observer' from its own perspective; but, on the other hand, {\it not} all external systems can be treated as ``observers'' from the perspective of a given observer. Instead of blankly stating that Quantum Mechanics does not allow observers to reason about other systems as observers, a more fine-grained solution is indeed possible. Being an observer is itself a relational property, one that is relative to another observer (or to a community of
other observers) and a background protocol (that restricts the timeframe and the possible histories). This doesn't make observers ``less real'' and less interesting, and it doesn't make Epistemic Logic less useful for reasoning about quantum scenarios, on the contrary.\footnote{In this sense, see some of our previous work \cite{BaSm08b,BaSm15,BaSm17,BS1,BS2}, on using epistemic logic to reason both about the informational properties of quantum systems (and in particular about eistemic characterizations of entanglement), and
about the epistemic states of classical observers in quantum environments.}


\begin{thebibliography}{9}
\bibitem{BaSm08b}	Baltag, A.; Smets, S.: A dynamic-logical perspective on quantum behavior, Studia Logica, 89:185-209 (2008)
\bibitem{BaSm15}	Baltag, A.; Smets, S.: Logics of Informational Interactions, Journal of Philosophical Logic, 44(6):595-607 (2015)
\bibitem{BaSm17}	Baltag, A.; Smets, S.: Modeling correlated information change: from conditional beliefs to quantum conditionals, Soft Computing, 21(6), 1523-1535 (2017)
\bibitem{BS2}  Baltag, A.; Smets, S.: Correlated Knowledge, An Epistemic-Logic View on Quantum Entanglement, International Journal of Theoretical Physics, 49:3005-3021 (2010)
\bibitem{BS1} Baltag, A.; Smets, S.: Correlated Information: A Logic for Multi- Partite Quantum Systems. In Coecke, B. and Panangaden, P. (eds.) Electronic Notes in Theoretical Computer Science ENTCS. Proceedings of the 6th Workshop on Quantum Physics and Logic, Oxford. vol. 270, pp.3-14 (2011)
\bibitem{Barwise} Barwise, J.; Perry, J.: Situations and Attitudes, Bradford Books, MIT Press (1983)
\bibitem{Boge} Boge, F.: Quantum information vs. epistemic logic: An analysis of the Frauchiger-Renner theorem. Foundations of Physics,   49:1143-1165, arXiv:1909.11889v1 (2019)
 \bibitem{Brukner1} Brukner, C.: On the quantum measurement problem, in R. Bertlmann and A. Zeilinger (eds): Quantum [Un]speakables II, Half a Century of Bell's Theorem. Springer (2017), arXiv:1507.05255v1 (2015)
\bibitem{Brukner} Brukner, C.: A no-go theorem for observer-independent facts, 	Entropy 20, 350 (2018)
\bibitem{Brukner2} Gu\'erin, P.A.; Baumann, V.; Del Santo, F.; Brukner, C.: A no-go theorem for the persistent reality of Wigner's friend's perception. Commun Phys 4, 93 (2021).
\bibitem{DeBrota} DeBrota, J.B; Fuchs, C.A; Schack, R.: Respecting One's Fellow: QBism's Analysis of Wigner's Friend, Foundations of Physics 50:1859-1874 (2020)
\bibitem{CR} Di Biagio, A.; Rovelli, C.: Stable Facts, Relative Facts, Foundations of Physics  51:30 (2021)
\bibitem{Corti} Corti, A.; Fano, V.; Tarozzi G.,: A Logico-Epistemic Investigation of Frauchiger and Renner's Paradox, International Journal of Theoretical Physics 62:54, (2023)
\bibitem{De85}	Deutsch, D.: Quantum theory as a universal physical theory, International Journal of Theoretical Physics, 24, I (1985)
\bibitem{Dretske}   Dretske, F.: Knowledge and the Flow of Information, MIT Press (1981)
\bibitem{Jordan} Elouard, C.; Lewalle, P.; Manikandan, S.K; Rogers, S; Frank, A. and Jordan, A.N: Quantum erasing the memory of Wigner's friend. Quantum, 2021-07-08, volume 5, page 498, arXiv:2009.09905v4 (2021)
\bibitem{FrRe18}	Frauchiger, D.; Renner, R.: Quantum theory cannot consistently describe the use of itself, Nature Communications, 9(1):3711. arXiv:1604.07422 (2018)
\bibitem{Halpern}    Fagin, R.; Halpern, J.; Moses, Y. and Vardi, M.: Reasoning about Knowledge, MIT Press (1995)
\bibitem{Healey} Healey, R.: Quantum Theory and the Limits of Objectivity Richard, Foundations of Physics, 48:1568-1589, (2018)
\bibitem{LaudisaRovelli} Laudisa, F. and Rovelli, C.: Relational Quantum Mechanics, E. N. Zalta (ed.), The Stanford Encyclopedia of Philosophy, online (2021)
\bibitem{Laloe} Lalo\"e, F.: Can quantum mechanics be considered consistent? a discussion of Frauchiger and Renner's argument. ArXiv:1802.06396v3 (2018)
\bibitem{LF} Leifer, M.S.; Spekkens, R.W.: A Bayesian approach to compativility, improvement and pooling of quantum states, 
    Journal of Physics A: Mathematical and Theorical, 47, 275301, arXiv:1110.1085v1 (2014)
\bibitem{Nielsen} Nielsen, M. A.; Chuang, I.L.: Quantum Information and Quantum Computation. Massachusetts Institute of Technology (2010)
\bibitem{NuRe21} 	Nurgalieva, N.; Renner, R.: Testing Quantum Theory with Thought Experiments, Contemporary Physics, 61(3):193-216,  arXiv:2106.05314v1 (2020)
\bibitem{NdR} Nurgalieva, N.; del Rio, L.: Inadequacy of Modal Logic in Quantum Settings, EPTCS 287:267-297 (2019)
\bibitem{Rovelli} Rovelli, C.: Relational Quantum Mechanics, International Journal of Theoretical Physics, 35(8):1637-1678, (1996)
\bibitem{RovelliNew} Adlam, E.; Rovelli, C. :Information is Physical: Cross-Perspective Links in Relational Quantum Mechanics, in Philosophy of Physics 1(1): 4, 1-19 (2023)
\bibitem{Shimony} Shimony, A.: Role of the Observer in Quantum Theory, American Journal of Physics, 31:755 (1963)
\bibitem{vDiHoKo07} van Ditmarsch, H.P; van der Hoek, W.; Kooi, B.: Dynamic Epistemic Logic, Springer (2007)
\bibitem{Wi61}	Wigner, E.P.:  Remarks on the mind-body question, in I.J. Good, The Scientist Speculates, London Heinemann (1961)
\bibitem{WvN} Waaijer, M.; van Neerven, J.: Relational Analysis of the Frauchiger-Renner Paradox and Interaction-Free Detection of Records from the Past, Foundations of Physics (2021)
\bibitem{Crull} Crull, E. and Bacciagaluppi, G.: Translation of W. Heisenberg: ``Ist eine deterministische Erg\"anzung der Quantenmechanik m\"oglich?'', Pittsburgh Phil-Sci Archive, (2011)
\bibitem{VN} von Neumann, J.: Mathematical Foundations of Quantum Mechanics , N. Wheeler (ed.) of the English translation, New Edition, 2018, Princeton University Press. First published in German in (1932)
\end{thebibliography}
\end{document}